\documentclass[12pt]{cernart}
\usepackage{epsf,epsfig,axodraw}

\newcommand{\sla}{\kern -5.4pt /}
\newcommand{\Dir}{\kern -6.4pt\Big{/}}
\newcommand{\Dirin}{\kern -10.4pt\Big{/}\kern 4.4pt}
\newcommand{\DDir}{\kern -7.6pt\Big{/}}
\newcommand{\DGir}{\kern -6.0pt\Big{/}}

\newcommand{\be}{\begin{equation}}
\newcommand{\ee}{\end{equation}}
\newcommand{\bea}{\begin{eqnarray}}
\newcommand{\eea}{\end{eqnarray}}
\newcommand{\beanon}{\begin{eqnarray*}}
\newcommand{\eeanon}{\end{eqnarray*}}
\newcommand{\ba}{\begin{array}}
\newcommand{\ea}{\end{array}}
\newcommand{\bd}{\begin{description}}
\newcommand{\ed}{\end{description}}
\newcommand{\bi}{\begin{itemize}}
\newcommand{\ei}{\end{itemize}}
\newcommand{\ben}{\begin{enumerate}}
\newcommand{\een}{\end{enumerate}}
\newcommand{\bc}{\begin{center}}
\newcommand{\ec}{\end{center}}

\newcommand{\bqas}{\begin{eqnarray*}}
\newcommand{\eqas}{\end{eqnarray*}}

\newcommand{\zb}{Z}
\newcommand{\wb}{W}

\newcommand{\bard}{\overline d}

\newcommand{\barnu}{\overline{\nu}}
\newcommand{\eqn}[1]{Eq.(\ref{#1})}

\newcommand{\fig}[1]{fig.~\ref{#1}}

\newcommand{\ar}{\rightarrow}
\newcommand{\parno}{\par\noindent}
\newcommand{\vsk}{\vskip 10 pt\noindent}

\newcommand{\wph}{{\tt WPHACT }}
\newcommand{\wphnosp}{{\tt WPHACT}}
\newcommand{\wphone}{{\tt WPHACT 1.0 }}
\newcommand{\wphonenosp}{{\tt WPHACT 1.0}}
\newcommand{\wphtwo}{{\tt WPHACT 2.0 }}
\newcommand{\wphtwonosp}{{\tt WPHACT 2.0}}

\newcommand{\pyt}{{\tt PYTHIA }}
\newcommand{\pytnosp}{{\tt PYTHIA}}
\newcommand{\veg}{{\tt VEGAS }}
\newcommand{\vegnosp}{{\tt VEGAS}}
\newcommand{\QEDPS}{{\tt QEDPS}}

\newcommand{\infi}{{\sl input file}}
\newcommand{\infis}{{\sl input files}}
\newcommand{\Infis}{{\sl Input files}}
\newcommand{\inca}{{\sl input card}}
\newcommand{\incas}{{\sl input cards}}

\def\pl #1 #2 #3 {{\it Phys.~Lett.} {\bf#1} (#2) #3}   
\def\np #1 #2 #3 {{\it Nucl.~Phys.} {\bf#1} (#2) #3}
\def\zp #1 #2 #3 {{\it Z.~Phys.} {\bf#1} (#2) #3}
\def\pr #1 #2 #3 {{\it Phys.~Rev.} {\bf#1} (#2) #3}
\def\prep #1 #2 #3 {{\it Phys.~Rep.} {\bf#1} (#2) #3}
\def\prl #1 #2 #3 {{\it Phys.~Rev.~Lett.} {\bf#1} (#2) #3}
\def\intj #1 #2 #3 {{\it Int. J. Mod. Phys.} {\bf#1} (#2) #3}
\def\mpl #1 #2 #3 {{\it Mod.~Phys.~Lett.} {\bf#1} (#2) #3}
\def\rmp #1 #2 #3 {{\it Rev. Mod. Phys.} {\bf#1} (#2) #3}
\def\cpc #1 #2 #3 {{\it Comp. Phys. Commun.} {\bf#1} (#2) #3}
\def\epj #1 #2 #3 {{\it Eur. Phys. J.} {\bf#1} (#2) #3}
\def\xx #1 #2 #3 {{\bf#1}, (#2) #3}

\begin{document}
\tolerance=100000
\thispagestyle{empty}
\setcounter{page}{0}
\bc{\large EUROPEAN ORGANIZATION FOR NUCLEAR RESEARCH} \ec

\vskip 1cm
\begin{flushright}
{\large CERN-EP/2002-037}\\
{\large DFTT 30/01}\\
{\large  PSI-PR-02-03}\\
{\rm 23 May 2002 \hspace*{.5 truecm}}\\
\end{flushright}

\vspace*{\fill}

\bc
{\Large \bf WPHACT 2.0 \\
\vskip 1truecm
A fully massive Monte Carlo generator for four
fermion physics at $e^+ e^-$ colliders.
\footnote{ Work supported by the European Union under contract 
HPRN-CT-2000-00149, by Ministero
dell' Universit\`a e della Ricerca Scientifica and by 
the Swiss Bundesamt f\"ur Bildung und Wissenschaft.\\[2 mm]
e-mail: ballestrero@to.infn.it,maina@to.infn.it,Elena.Accomando@psi.ch}}\\[2.cm]
{\large Elena Accomando$^1$, Alessandro Ballestrero$^{2,3}$ and 
    Ezio Maina$^{4,3}$}\\[.3 cm]
{\it $^1$ Paul Scherrer Institut, CH-5232 Villigen-PSI, Switzerland}\\
{\it $^2$CERN, EP Division}\\
{\it $^3$INFN, Sezione di Torino, Italy}\\
{\it $^4$Dipartimento di Fisica Teorica, Universit\`a di Torino, Italy}\\
\ec

\vspace*{\fill}

\begin{abstract}
{\normalsize
\noindent
\wphtwo is the new fully massive version of a MC program and
unweighted event generator 
which computes all Standard Model processes with four fermions in the final 
state at $e^+ e^-$ colliders. 
The program can now generate unweighted events for any subset of 
all four fermion final 
states in  a single run, by making use of dedicated pre-samples which can
cover the entire phase space.
Improvements with respect to \wphone include the Imaginary Fermion Loop 
gauge restoring scheme, new phase space
mappings, a new input system, the possibility to  compute subsets 
of Feynman diagrams and options for including ISR via \QEDPS , running 
$\alpha_{QED}$, CKM mixing, resonances in $q \bar q$ channels.  

}
\end{abstract}

\vspace*{\fill}

\section*{NEW VERSION SUMMARY}
\vspace{10pt}

\noindent {\sl Title of program:} {\tt WPHACT}, version {\tt 2.0}. 
\vskip 0.2cm
\noindent{\sl Program obtainable from:} CPC Program Library, Queen's University
of Belfast, N.~Ireland. Also available in 
{\tt http://www.to.infn.it/$^\sim$ballestr/}
\vskip 0.2cm
\noindent{\sl Reference to previous version:} \cpc 99 1997 270
\vskip 0.2cm
\noindent{\sl Authors of original program:} E. Accomando and A. Ballestrero.
\vskip 0.2cm
\noindent {\sl Catalogue identifier of previous version:} ADEN 
\vskip 0.2cm
\noindent {\sl The new version supersedes the previous one }
\vskip 0.2cm
\noindent{\sl Computer:} Any computer with {\tt FORTRAN77} compiler which admits
structures. \wph has been tested on Compaq ALPHA and HP stations 
as well as on Linux Intel PC. For this last type of machines the Portland 
Group pgf77 
compiler has been employed, while the GNU compiler cannot be used as it does 
not have the {\tt STRUCTURE} extension. 
Compaq and Portland Group {\tt FORTRAN90} compilers can also be used.  
\vskip 0.2cm
\noindent{\sl Operating systems:} UNIX, LINUX, VMS.  
\vskip 0.2cm
\noindent{{\sl Programming language used:} FORTRAN 77}
\vskip 0.2cm
\noindent{{\sl Memory required to execute with typical data:} $\approx$ 
500 KByte}
\vskip 0.2cm
\noindent{{\sl No. of bits in a word: } 32}
\vskip 0.2cm
\noindent{\sl No. of processors used:} one
\vskip 0.2cm
\noindent{{\sl No. of bytes in distributed program:} $\approx$ 2 Mb }
\vskip 0.2cm
\noindent{\sl Distribution format:} tar gzip file ($\approx$ 700 Kb 
including a version of {\tt PYTHIA})
\vskip 0.2cm
\noindent{\sl Additional keywords:} massive matrix elements, gauge invariance, 
fermion loop, single W, single Z, running $\alpha_{QED}$, CKM mixing, 
resonances. 

\vskip 0.2cm

\noindent{\sl Nature of physical problem}: 
All $e^+e^-$ Standard Model (SM) processes with four fermion final state
are necessary for detailed studies of the properties
of the W and of the Higgs bosons, for measuring double and 
single boson production,
for high precision tests of the SM and for evaluating backgrounds to searches.
One wants therefore a Monte Carlo event generator which provides 
an accurate description of  all four fermion processes in the largest possible 
region of phase space and which can be used for realistic experimental 
simulations.

\vskip 0.2cm
\noindent {\sl Method of solution}: 
Full tree level matrix elements for all  processes are computed
by means of subroutines which make use of the helicity formalism of ref.
\cite{method}-\cite{phact}. The speed in computing these amplitudes, that 
the above mentioned method allows, is  essential to take exactly into
account fermion masses and to obtain high precision in a reasonable amount of
CPU time also in regions with large and possibly overlapping enhancements. 
\vskip 0.2cm

\noindent {\sl Reasons for the new version:}  
Extend the code to cover in a reliable way all $t$-channel dominated processes 
and low $f\bar f$ invariant mass regions. 
Introduce the possibility to simultaneously generate unweighted events
for all processes or any user-selected subset. 
Include photon $p_t$ and CKM mixing.
\vskip 0.2cm

\noindent {\sl Summary of revisions:}
Fully massive matrix elements, new unweighted event generation for 
any subset of  four fermion final 
states in  a single run, implementation of the Imaginary Fermion Loop 
gauge restoring scheme, new phase space mappings, new input system, 
possibility to  compute subsets of Feynman diagrams and options for 
including ISR via \QEDPS , running $\alpha_{QED}$, CKM mixing, 
resonances in $q \bar q$ channels.  
\vskip 0.2cm

\vskip 0.2cm
\noindent{{\sl Restrictions on the complexity of the problem} }
\noindent
QED radiative corrections are implemented 
only in Initial State Radiation and in Coulomb corrections.
Final state radiation is not computed  when 
unweighted events are not requested.
Only the imaginary part of Fermion Loop corrections is implemented.
QCD corrections are introduced in an approximate way.
\vskip 0.2cm
\noindent{\sl Typical running time:}
The running time strongly depends on the process considered and on the
precision requested. 
The example reported in the test run took about 90 minutes on a 500 MHz ALPHA
computer, which corresponds to effectively generating more than 1500
weighted $e^-e^+ \mu^-\mu^+$ events per second.
\noindent

\vskip 0.2cm

\noindent{{\sl Unusual features of the program:} }
\noindent
{\tt STRUCTURE} declarations are used.
\eject
\section*{Long Write-Up}
\vsk

\section{ Introduction }

Four fermion final states in $e^+ e^-$ collisions, which involve electroweak
boson pair production, are of special interest since they
allow the mechanism of spontaneous symmetry 
breaking and the non abelian structure of the Standard Model (SM) 
 to be directly tested 
by the experiments. Moreover they provide a very important background to
most searches for new physics.

LEP2 has provided in this respect an ideal testing ground for the SM.
After the end of the period of data taking, the Collaborations are working
towards the final results and combinations.
W properties have been measured with great accuracy.  New bounds on 
anomalous trilinear  gauge boson couplings and new limits on the Higgs mass  
and Susy particles have been set. Single W, single Z, ZZ and Z$\gamma ^*$ cross 
sections have been determined for the first time.   
Hopefully these studies will be continued with much higher statistics and
energy at a future $e^+e^-$ Linear Collider.

In parallel  and in strict collaboration with the experimental activity, a
considerable theoretical effort has been devoted to the analysis of four
fermion physics and to the production and improvement of dedicated Monte 
Carlos.  Much of this work is  documented in the two yellow reports of the 
LEP2 CERN Workshops~\cite{yr,yr2k} . 

A first version of {\tt WPHACT}, a Monte Carlo program and
unweighted event generator for all processes $e^+e^-\rightarrow 4f$ was 
published \cite{wph1} at the beginning of LEP2 activity.
It was mainly aimed at describing W pair production
and Higgs physics, but it allowed to compute any four fermion final state. 
\wphone was extensively compared 
with several other codes \cite{fort}
during the first LEP2 workshop~\cite{wweg,dpeg},
and demonstrated its reliability.
All these early codes, with the exception of {\tt grc4f}\cite{grc4f},
were based on massless matrix elements. \wphone took exactly into account
only the b mass. The reason for such a choice was essentially the huge amount 
of CPU time needed for massive calculations and the very good results
given by the massless approximation to all  WW, ZZ and Higgs physics.  
As a consequence, however, the codes were
inadequate to describe processes dominated by $t$-channel exchanges
like single W production ($e^+e^-\rightarrow e^-\bar\nu_e f\bar f'$ at
small electron scattering angle), single Z   and $\gamma \gamma$ like 
configurations ($e^+e^-\rightarrow e^- e^+  f\bar f$ in which one or both
electrons are  undetected). 

After a few years of data taking it has become evident the necessity of
having an accurate description of such regions, and in particular of single
W production which is relevant for triple gauge boson 
coupling measurements and as a background to New Physics searches. 
Small angle electron scattering
is singular in the massless limit and this makes it extremely sensitive to
violations of U(1) gauge invariance \cite{u1gi} like those produced by a naive
introduction of the decay width in W propagators.   It is therefore mandatory 
to introduce fermion masses and to adopt a consistent gauge restoring scheme. 
For such a reason a new fully massive version of {\tt WPHACT} was created
and the Fermion Loop method~\cite{bhf} of restoring gauge invariance was 
extended to the massive case~\cite{ifl,efl}.

Many fully massive $4f$ MC programs are now available.
\wphnosp, {\tt grc4f}\cite{grc4f}, {\tt KORALW}\cite{koralw} and 
{\tt COMPHEP}\cite{comphep} are based on Feynman diagrams\footnote{
{\tt grc4f} and {\tt KORALW} adopt the same matrix elements based on
{\tt GRACE}\cite{grace}} while {\tt NEXTCALIBUR}\cite{nextcalibur}
and {\tt SWAP}\cite{swap} are based on the Dyson--Schwinger equations, 
a set of recursive relations among Green functions.
Comparisons between the different codes have shown good technical
agreement in several
benchmark processes, as recently summarized in \cite{4f}.

Other advances of great significance documented in \cite{4f} regard
higher order corrections to four fermion  WW processes in
the so called double pole approximation (DPA)~\cite{dpa,racoonww,yfsww}. 
This feature, which is not implemented in \wphnosp ,
has led to a major improvement in the theoretical uncertainty for WW cross
sections and distributions, which has decreased from the previous estimate 
of 2\% , which was comparable with the experimental errors, 
to  about 0.5\% . 
 
In the present paper we present a full description and  documentation of the 
new version of \wph in which many new features and improvements have been
added. 

In order to meet some of the additional requirements put forward by the ever
increasing accuracy of experimental measurements
at present and future accelerators we have included:
\begin{itemize}
\item fully massive matrix elements and phase spaces for all $4f$ final states
\item new phase space mappings for low mass and small scattering angle regions
\item Fermion Loop corrections in the IFL scheme \cite{ifl}
\item explicit generation of ISR photons with $p_\perp$ effects via an 
interface to \QEDPS \cite{qedps} which can be used as an alternative to
the previous Structure Function (SF) approach
\item possibility to use running  $\alpha_{QED}$ for $t$-channel or low invariant
masses 
\item support of CKM mixing
\item resonances in  $q \bar q$ channels and new low mass hadronization
with an interface to the routines of Ref.~\cite{boonek} 
\item possibility to compute relevant subsets of Feynman diagrams 
\item optional beamstrahlung effects for future Linear Colliders generated 
with the help of {\tt circe} routines~\cite{circe}   
\end{itemize}

Since generation of large samples of fully simulated events is best performed
if all processes or specific classes of them  can be generated simultaneously, 
we have introduced the possibility of 
\begin{itemize}
\item unweighted events generation for any subset of massive $4f$ final states
      in a single run
\end{itemize}

We have moreover implemented new routines which provide 
\begin{itemize}
\item a simplified and more user friendly procedure for specifying the 
      program input. 
\end{itemize}

These new features and the  full set of available input settings is 
described in the following. One example of test run is given at the end.
 
\section{ New and improved features}
We have retained all the best features of \wphonenosp. 
One of the main strengths of the code is the use of the helicity 
amplitude method described in Ref.~\cite{method}.
The code for \wph amplitudes has 
been completely written with the help of {\tt PHACT} \cite{phact} 
({\bf P}rogram for {\bf H}elicity {\bf A}mplitudes {\bf C}alculations with 
{\bf T}au matrices).
With this formalism it is possible to evaluate 
tree-level matrix elements in a very fast and efficient way by means of a 
modular scheme which
stores for later use subdiagrams of increasing size and complexity. 
Moreover the massive case is a rather straightforward extension
of the massless one.
It is in fact based on the same modular and diagrammatic approach and all 
helicity
combinations are computed simultaneously. As a consequence, the 
code with massive amplitudes written in this way is only about five times
slower than the previous massless one. This is to be compared with the fact
that the number of helicity states increases by a factor of eight and new
terms in the diagram evaluation appear.  Moreover, since 1996 the speed of
available computers has increased by a factor of about thirty and massive
calculations are presently faster than massless ones in the old days. 
In any case, since massless amplitudes provide an excellent approximation to 
the full result in all cases which do not exhibit collinear or mass 
singularities,
we have maintained the option to use the faster massless amplitudes. 
The fast evaluation of matrix elements, both massive and massless,
combined with the use of the adaptive integration 
routine \veg \cite{vegas}, is particularly useful when producing 
distributions at
parton level with very high precision in each bin, or when generating complete 
and statistically significant samples of unweighted events. 
As in \wphonenosp, any distribution at parton level and any event sample can be 
produced 
while evaluating the cross section.

\subsection{ Processes }

\wph computes all SM processes with four fermions in the final state at 
$e^+ e^-$ colliders. Final states with $t$ quarks are not considered, as 
the $t$'s are known to decay immediately to 3 other fermions.

\par

The list of available processes is given in Tables 1, 2 and 3. 
All possible channels are divided in four classes. 
The first (CC) and the third (NC) one contain all so called 
charged  and neutral current processes respectively.  
The second (MIX) is the class in which both neutral and charged current
contributions are present.  
These three classes exhaust the four fermion processes. The corresponding matrix
elements can be computed with all fermion masses taken into account or in
the massless approximation. 
Contributions from Higgs diagrams are not available in these first three
classes.
The last class in Table~3 (NC+Higgs) contains again all neutral current 
processes with $b$'s, $c$'s and $\tau$'s. These are specific for Higgs
studies: requesting these last processes 
(flag {\tt iproc} ranging from 33 to 53) one 
can  compute the Higgs signal  or the background or the complete 
Higgs+background+interference. In this case the first seven ones (33 to 39
included) are intended for Higgs decaying to $b\bar b$, the second group of 
seven  for $H\ar c\bar c$ and the last group  for $H\ar \tau^+\tau^- $. 
The corresponding matrix elements are evaluated with massive $b$'s, $c$'s and 
$\tau$'s respectively, with all other particles taken as massless both in the 
matrix elements and phase space.   

With respect to the previous version of \wphnosp, new channels have been added.
The NC table 
has been extended by including processes with $b$'s in the final 
state, while the NC+Higgs table 
now contains also channels in which the Higgs 
boson decays into $c$'s and $\tau$'s.

We have introduced the possibility to take into account CKM mixing,
which  provides  a more accurate description of the flavour content
of  hadronic final states. This might be relevant in 
evaluating backgrounds to Higgs and new particle searches.  
In the CC class we have included all additional channels which
are induced by $d \leftrightarrow s$ and $s \leftrightarrow b$ mixing.
The processes which are induced by $d \leftrightarrow b$, which are 
much more strongly suppressed, have been neglected, as the
multiperipheral type contributions to NC processes
 $e^+e^-\ar \nu_e \bar \nu_e q \bar q$, in which two W's attach to a single 
 fermion line and one sums on the flavour of the virtual quark.
In the NC class, the contributions due to CKM mixing are doubly suppressed.
 Nonetheless, we have included the CC exchange contributions
to the three processes
     $$
                    e^+ e^- \ar   s  \bar u  u  \bar s,\  
                    d  \bar c  c  \bar d,\
                    b  \bar c  c  \bar b
     $$
 which can be produced when CKM mixing is turned on.

In \wph the momenta of the final state partons are
called {\tt p3}, {\tt p4}, {\tt p5}, {\tt p6}. They are assigned 
in the order in which the particles appear in  Tables 1, 2 and 3.
Therefore, for instance, the first process in Table 1 reads:
$$
e^+(p_1) e^-(p_2) \ar \mu^-(p_3) \bar\nu_\mu(p_4) \nu_\tau (p_5)\tau^+(p_6)
$$
The order of the  
particles is important when one wants to impose cuts or to compute distributions
at parton level.

It has to be noticed however that for CC and Mixed processes the order of the 
particles in \wph is different from the one used in 
\pytnosp ~\cite{pythia}.
We conform to the convention of \pyt when in \wph the  
momenta are passed to the high energy standard 
{\tt COMMON/HEPEVT}~\cite{tb}.
Moreover, when 
\pyt is called, all 3-momenta are reversed in order to conform  
to the convention  for which $e^-$ is in the positive $z$ direction 
(\wph default is that the incoming $e^+$ is in the $+z$ direction).
\par

Many different final states give the same cross section at parton level up
to mass effects. This is the case for instance of  
$\mu^-$  $\mu^+$  $d$  $\bar d$ and $\tau^-$  $\tau^+$  $s$  $\bar s$ if 
mass differences are neglected.
For this reason each final state is described by a combination of two flags,
 {\tt iproc} 
and {\tt ich}, which unambiguously identify it. The first flag refers to a
group of processes which are described by the same set of diagrams, 
the second to the specific final state in the group.
Processes which differ by a final state CKM rotation are also distinguished by
 {\tt ich}.
It has to be 
noticed that for CC processes charge conjugate final states belong to the 
same {\tt iproc}. The amplitudes for a CC final state and its charge 
conjugate, for a given set of four momenta, are not identical. They are related
one another by a final state parity transformation.
This implies that if  cuts are not invariant under $P$, the two cross 
sections will be different. In \wph they are
considered separately and are identified by different values of {\tt ich}.
When identical particles appear in the final state the cuts must obviously
be symmetrical with respect to their exchange. This is not checked for on input
and unphysical non-symmetric cuts can produce unreliable results.

\par

Since our approach is based on Feynman diagrams, it is possible to compute
subsets of the diagrams for a given amplitude.
While this procedure is not in general gauge--invariant and should as a
consequence be used with extreme care, it is nonetheless extremely useful
in practice in order to compare and combine the results of several experiments 
which adopt different selection strategies. In effect, when appropriate 
cuts evidentiate quasi on shell contributions of intermediate vector bosons,
the subset of diagrams which would correspond to the production times decay 
approximation gives by far the dominant contribution to the cross section.
The most famous example is the so called CC03 contribution to WW production,
but one can of course consider also ZZ or Z$\gamma*$.
$t$-channel dominated processes like single Z or single W also belong to this
category, as one may consider that the photon in $t$-channel is almost on
shell (Weiszacker-Williams approximation). 
Signal definition in terms of a subset
of diagrams is also useful when one can take advantage of a specialized program
which describes phenomenologically a set of processes, for instance
gamma--gamma physics, which are known to have important non--perturbative
contributions. In other cases it has been successfully employed in order
to take into account higher order corrections which were available only for
the dominant part of an amplitude.
In sections~\ref{subset}  and \ref{higgs} we describe the subsets
of diagrams which can be selected in \wphtwonosp.

\par

All neutral current processes are normally computed at order $ \alpha ^4$.
When there are four quarks in the final state, there are however 
contributions $O(\alpha ^2 \alpha_s ^2)$ from diagrams with gluon   
exchange. These contributions are enhanced by the ratio of the
strong to the electroweak coupling and can be relevant for some processes,
energies and cuts. 
However, these contributions 
are already, at least partially,
accounted for by two quark final states when parton 
showering and hadronization are performed. In order to avoid a problematic double 
counting, in \wphtwo gluon exchange diagrams have been switched off by default.
The corresponding flag is {\tt iqu} which is set to zero in a  {\tt DATA}
statement. Setting {\tt iqu=1} enables the computation of these QCD
diagrams.
  
\subsection {Phase space and integration}\label{ph}

Most processes which have been studied in the first part of LEP2 activity
are dominated by the production of boson pairs (i.e. WW, ZZ, Z$h$ or $hA$).
In this case, the four final fermions can be conveniently 
divided into two sets corresponding to possible decays of a W in CC
processes, or to possible decays of a Z or Higgs boson in NC contributions.
The cross section is strongly enhanced when the two fermion--pair invariant
masses $M_1$ and $M_2$ are close to the intermediate boson masses.
Appropriate phase space mappings were already available in \wphonenosp. 

In the new version of \wph new mappings dedicated to cover both low invariant
mass and small scattering angle regions have been implemented.

In NC processes an additional peak for low invariant masses,
due to the photon $s$--channel propagator, may be present.
In order to treat all possible kinematical configurations of this kind, a 
convenient double mapping which accounts for both Z and $\gamma^*$
peaks on the same variable $M_i$ (i=1 or 2)   is performed.

When both the $M_1$ and $M_2$ invariant mass peaks 
are mapped into a uniform distribution we refer to the
corresponding phase space as double resonant. If only one or no 
variable transformation is performed, we have respectively a single resonant 
or a non resonant mapping. 

In \wph there is the possibility to choose among these mappings as explained
in Sec.~\ref{phasespace}.
For Mixed (or NC+Higgs) processes, the peaking structures of CC (or Higgs) 
and NC contributions are different. \wph by default integrates 
separately the two contributions, adding the interference to one of them.
The program uses two different phase spaces for the two
different parts of the calculation.
When generating events, the interference is treated in such a way to avoid
negative weights.

New mappings have also been introduced to deal with forward scattering.
The processes in which one or two electrons (or positrons)  
are emitted at very small angle with respect to the beam
present collinear enhancements which are logarithmic in the ratio of the
electron mass to the total collider energy.
They are therefore extremely sensitive to numerical
cancellations which, if not properly treated, can lead to strong numerical
instabilities. Our approach to such integration is based
on the method of Ref.~\cite{swphasespace} and can be switched on by the flag 
{\tt ismallangle} as it will be described in Sec.~\ref{phasespace}.
In this case the user can choose among single or non resonant
mappings for the invariant mass of the fermion pair which contains no
forward electrons (or positrons). 
Even with the new tools, if cuts are such that one gets important 
contributions from extreme regions, it is safer to run \wph in 
quadruple precision (with obvious modifications to the code),
  or to test that {\tt real*8} result agrees with {\tt real*16}. 
An example of these possible dangerous contributions is given, for processes
with outgoing electrons at LEP2, by the region in which $|q_\gamma^2|$, 
the absolute value of the squared momentum of the exchanged t-channel photon,
becomes much lower than 
$10^{-14}$. In such cases, also the calculation in double precision of  
$|q_\gamma^2|$  from the external four-momenta becomes unreliable.

All integrals are computed with {\tt VEGAS}. With this routine
it is preferable to use more than one iteration since between iterations
{\tt VEGAS} adapts its phase space grid to the integrand at hand in an
effort to decrease the overall variance. In each iteration
the integral is evaluated and in the end the various results are combined 
together. This allows to optimize the number of points evaluated in each
region of the integration variables. It may also be
useful to perform some thermalizing iterations with  a smaller number
of points. The thermalizing iterations are used only
to adjust the grid and not for the final result. 

The adaptivity of {\tt VEGAS} is such that, even if most cuts are  
implemented in the program with the use of {\tt if} statements
which act as $\theta$ functions, usually this does not
correspond to a substantial loss in time and precision.
The cuts on the invariant masses which are function of integration  variables
are implemented directly as integration limits. 

\subsection {Radiative effects}

In view of the experimental accuracy reached at LEP2, which will be 
further increased at future Linear Colliders, the inclusion of
QED higher order radiative corrections is an unavoidable step for accurate
theoretical predictions of cross sections and distributions.
 
The exact calculation of $O(\alpha )$ electroweak corrections for four 
fermion processes is very difficult and at present available  only 
for the WW signal in Double Pole Approximation (DPA) \cite{dpa} and 
implemented exclusively in
{\tt RACOONWW} \cite{racoonww} and {\tt YFSWW3} \cite{yfsww,koralw}.
 
In the rest of this section we will describe how important radiative
effects as 
Initial State Radiation (ISR), Final State Radiation (FSR),
Coulomb Corrections and Fermion Loops are accounted
for in \wphtwonosp. We will also describe the implementation of 
running $\alpha_{QED}$ which can give a rather good estimate of higher order
effects in $t$-channel or low $f\bar f$ mass regions. 

\subsubsection {Initial state radiation}\label{isr}
 
Initial state radiation  is known to give large corrections.
The Structure Function (SF) formalism, the Parton Shower (PS), and the Yennie, 
Frautschi and Suura algorithm (YFS), 
are the three most widely used methods to describe ISR in $e^+e^-$ 
annihilation processes. The last two approaches take into account the
transverse momentum ($p_\perp$) of the emitted photons. 
The SF formalism usually generates 
collinear radiation, but can include $p_\perp$ effects via
$p_\perp$--dependent SF.
 
In the previous version of the program, the relevant Leading Log (LL) QED 
radiative corrections due to ISR were implemented via the
Structure Function method according to the soft and collinear 
approximation of Ref.~\cite{isr1} ($p_\perp$ independent).

In the new version, we have extended the treatment of the QED radiation by 
incorporating the Monte Carlo program \QEDPS \cite{qedps}, based on the parton 
shower algorithm in QED, which solves the DGLAP equations in the LL 
approximation and generates photons $p_\perp$.
The transverse momentum of emitted photons, neglected in the collinear 
limit, can have sizable effects on the cross section and might 
modify the shape of the distributions. This is an important effect to take 
into account. However one must be somewhat careful because,
while in the collinear  and massless limit the electrons, after radiating, 
remain
on mass shell, this is no more true when the emitted photons have $p_\perp$ and
when electron masses are accounted for. In such cases the commonly used
approximation of forcing electrons to be on mass shell after ISR emission
may not be always reliable.

The user can select between the SF and PS options via the input flag {\tt isr}
(see Sec.~\ref{radiation}).

An important source of theoretical uncertainty related to radiative effects 
is the choice of the correct energy scale in forward scatterings, like single 
W, single Z and $\gamma\gamma$ processes. Commonly applied to annihilation 
processes, ISR turns out to require a more careful treatment for $t$--channel 
dominated scatterings, especially when including the photon $p_\perp$. 
The choice of the scale is determined by physics considerations and, 
when possible, by comparing with higher order results. 
The $natural$  scale used for $s$--channel processes, that is the total energy 
squared $s$, might not 
give a good description of the radiative emission, when 
applied to multi-scale processes dominated by the low transfered momentum 
$|q^2_\gamma |$ of the exchanged $t$-channel photon.
 
For small angle Bhabha scattering and multi-peripheral $\gamma\gamma$ 
processes, where exact QED radiative corrections are already known,
the full prediction and the result obtained by using SF at the scale 
$|q^2_\gamma |$ are in good agreement \cite{sfgg}. In all other
cases, where no exact calculation is available, it has been recently 
shown that a comparison with the soft limit of the $O(\alpha )$ corrections
can determine the scales to be used  
in SF and PS \cite{sfpv,sfkur}.  
Detailed analysis, concerning the sensitivity of $t$-channel scatterings to
the choice of the energy scale in the presence of ISR, have been 
performed \cite{sfpv,sfkur,nextcalibur}, leading to a better understanding of 
the behaviour of the radiation for single W and $\gamma\gamma$ processes .
The general finding is that the choice of the scale describing the SF 
evolution can produce differences up to $8\%$, and there is
a widespread consensus on the fact that the $s$ scale is not the most 
appropriate.
 
When non strictly collinear ISR is introduced by any of the methods,  
PS, YFS or $p_\perp$--dependent SF, the  situation is even more delicate.
In these approaches in fact, after radiative emission, the off-shell electrons 
are projected on mass-shell for computing the amplitude, as already mentioned.
This approximation is not valid for high $p_\perp$ ISR photons
and low  momentum transfer, $|q^2_\gamma |\simeq 0$, as can be easily 
realized  by
comparing with  the exact kinematic of the virtual 
$t$-channel photon.
Therefore, slightly different  $p_\perp$ spectra can lead to 
rather different  and sometimes inconsistent results in the extreme 
forward regions. However, the use of $|q^2_\gamma |$ or similar scale 
implies a high
suppression of the radiation in such regions. As a consequence, 
 when taking into account the $p_\perp$ of ISR photons, one can
avoid the above mentioned problem because the low 
momentum transfer events are almost never accompanied by high 
radiation and therefore by high  $p_\perp$.
In this sense, choosing an ISR radiation scale different from $s$ is 
unavoidable, and not just a
possible option as for the SF collinear case.

A more consistent theoretical study of the problem of radiation in 
multi-scale processes has been recently performed in Ref.~\cite{giampisr}, 
but it has not yet been implemented in a MC code.

In \wphtwonosp, ISR is treated  as follows. 
When ISR is included via SF ({\tt isr=1}), the energy scale relevant to the 
radiative emission is fixed to be $q^2=s$, independently of the process.
When ISR is implemented via PS ({\tt isr=2}), \wph follows instead a 
strategy in part similar to that of Ref.~\cite{sfkur}. 
If the final state does not contain 
any electrons or positrons, the energy scale is fixed to be $q^2=s$. For 
processes involving at least one electron (or positron),
the program generates, as a first step, 
the complete kinematics in the absence of
photon emission. The transfered momentum $|q^2_\gamma |$ of the photon 
exchanged in the $t$--channel is then computed. If 
$|q^2_\gamma |\le 10m_e^2$ ($m_e$ is the electron mass), the program 
evaluates the differential cross section without ISR.
In all other cases, \wphtwo 
goes back to the starting point, switches on photon emission via \QEDPS,
recomputes the kinematics and evaluates the differential cross section.
If $10m_e^2\le |q^2_\gamma |\le s/100$,  
the scale $q^2=|q^2_\gamma |$ is used, otherwise $q^2=s$.

\begin{figure}[htb]
  \begin{center}
  \unitlength 1cm
 \begin{picture}(15,13.5) (0,0)
\put(-2.3,-7.6){{\includegraphics{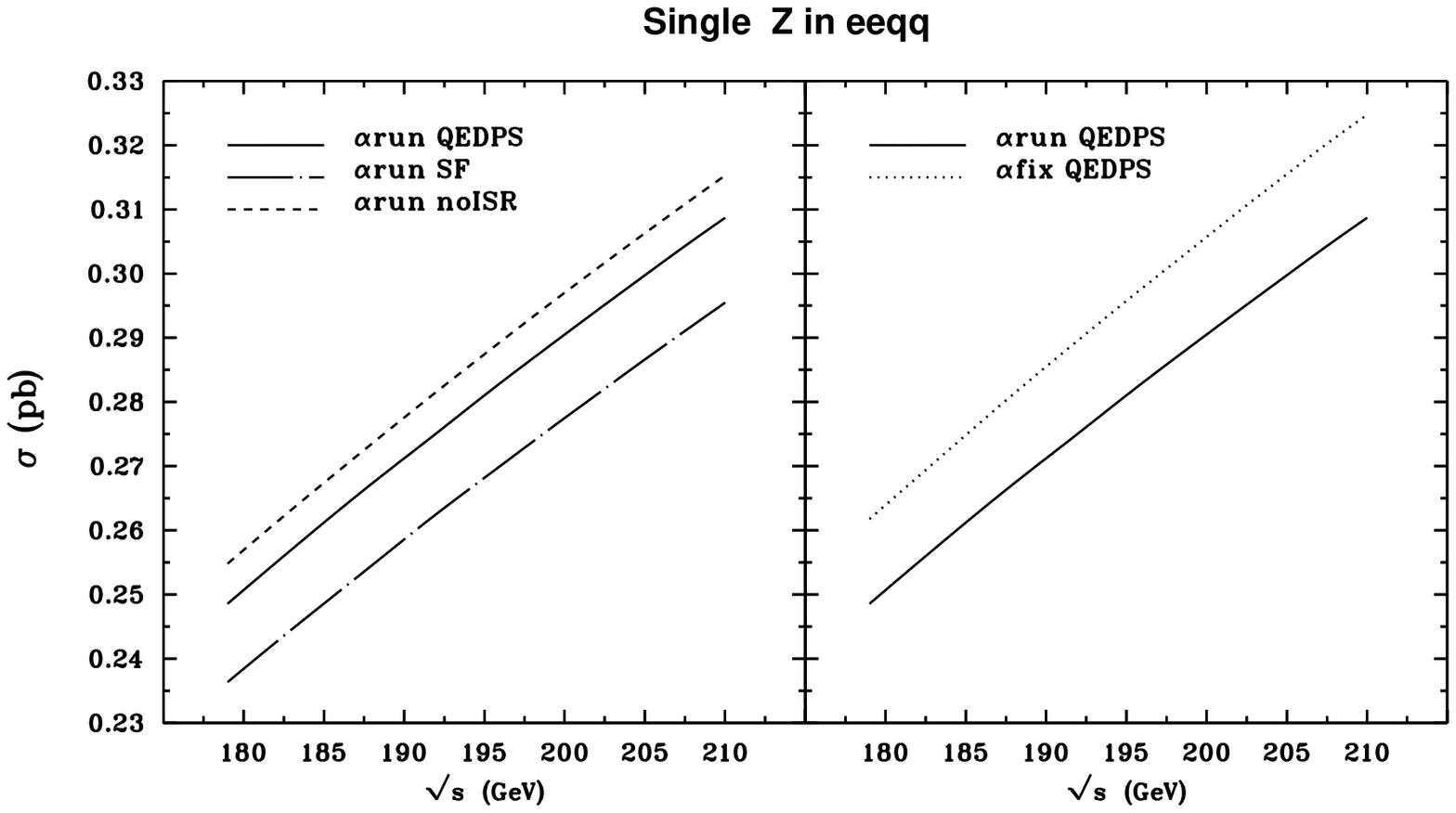}}}
\put(-2.3,-0.2){{\includegraphics{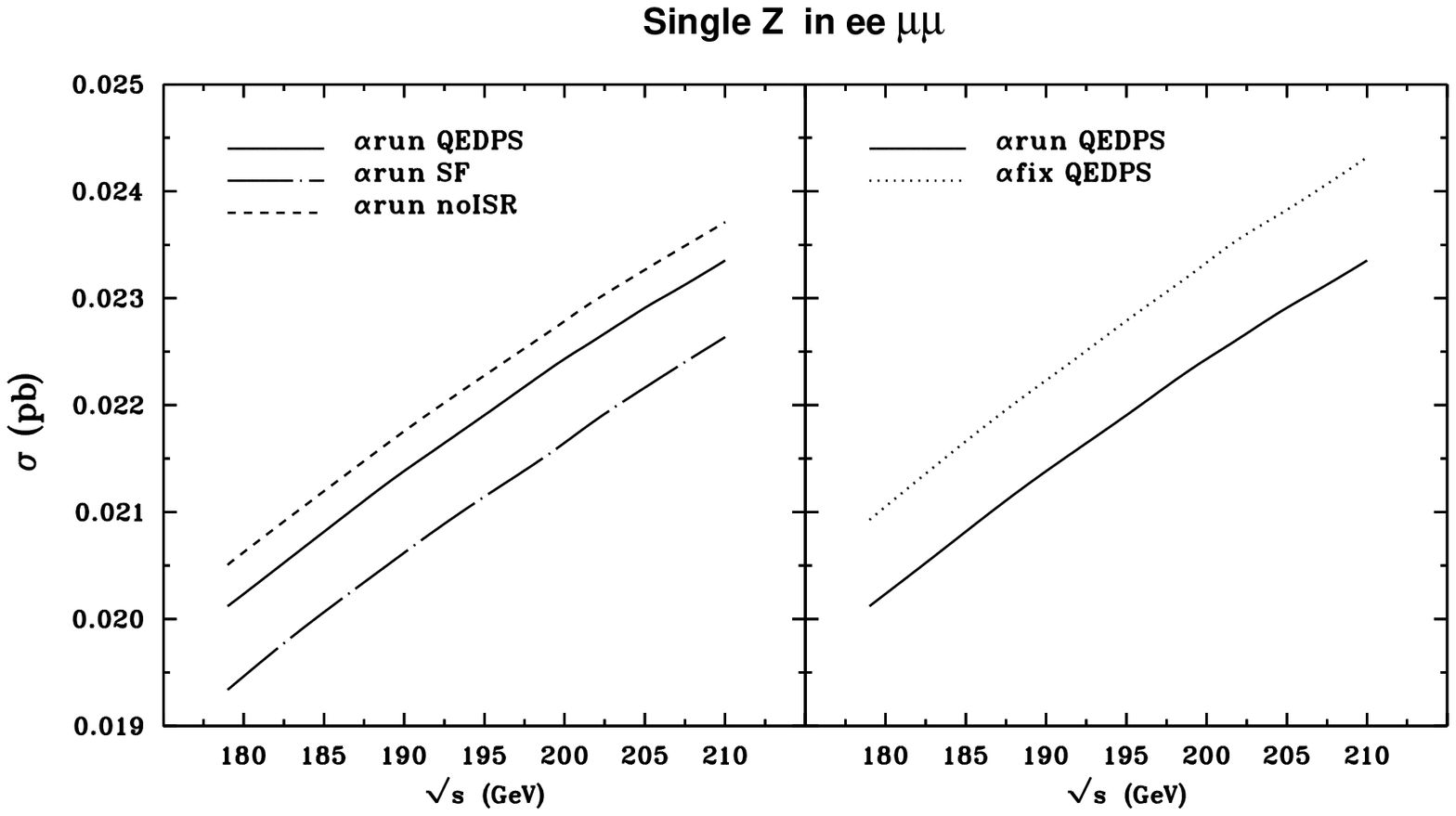}}}
\end{picture}
\caption[]{Cross sections for $e^-e^+\mu^-\mu^+$ (upper) and for the sum of
 $e^-e^+q\bar q$ ($q=u,d,c,s,b$).
Cuts for single Z region are defined in the text.
Left--hand plots compare different ISR implementations, right--hand ones show
the effect of running $\alpha_{QED}$.

}
\label{singleZq}
\ec
\end{figure}

In \fig {singleZq} some results regarding the so-called single Z processes
are reported. In the upper part the cross section for 
$e^-e^+\mu^-\mu^+$ is given as a function
of the energy. In the lower part we present the cross section for the sum of all
processes  $e^-e^+q\bar q$ ($q=u,d,c,s,b$). The following cuts have been 
applied: $m(f\bar f) > 60$ GeV
($f\neq e$), $e^-$ within 12 degrees from its forward direction,
 $e^+$ between 60 and 168 degrees (\wph convention is incoming 
$e^+$ at $0^\circ$). 
In the left hand side plots we compare different
treatments of ISR. From them one can see the numerical relevance of
{\tt QEDPS} as implemented in \wph with respect to SF with $s$-scale.
The {\tt QEDPS} curve falls in between the SF one and the curve 
in which no ISR effect is considered. For these set of cuts
the difference between SF ($s$-scale) and QEDPS ($t$-scale) is
of the order of 4\%.  All three curves have
been computed using running $\alpha _{em}$ coupling in photon vertices (see
\ref{r_alpha}).

\subsubsection {Final state radiation}\label{fsr}
   
Radiative corrections and photon emission in the final state are not 
directly computed in \wphnosp , unless unweighted events are generated.
In this case the variable {\tt irad} which is in a {\tt DATA} statement of
the routine {\tt wph.f}, determines how FSR is computed. If {\tt irad}=0,
no FSR photons are emitted. If {\tt irad}=1 and \pyt is called ({\tt
  ijetset=1}), FSR radiation is performed via \pyt itself. 
  Having final state radiation performed by an hadronization 
  package can be preferable in case of quarks in the final state, where 
  interplay between strong and electromagnetic corrections takes place.
  In the case of emission from leptons, one can instead prefer to have
  FSR from the {\tt PHOTOS}~\cite{photos} dedicated package. For purely
  leptonic processes,  this is achieved by  setting 
  {\tt irad}=2 and {\tt ijetset=1}. For semileptonic channels, the same
  set of flags produces {\tt PHOTOS} FSR for the leptonic pair
  and \pyt FSR for the quark pair.  

\subsubsection{Coulomb Correction}\label{coulomb}

The Coulomb singularity is relevant only for CC and Mixed processes near the 
W-pair threshold. However, in that kinematical region it gives a sizable 
correction of a few percent. In \wph the Coulomb correction can now be 
computed with 
four different methods. The input parameter {\tt icoul} (see Sec.~\ref{corrections}) 
allows one to choose among the different implementations here below 
summarized. We define the following quantities:

\be
p^2={1\over{4s}}(s^2-2s(s_1+s_2)+(s_1-s_2)^2);~~~~~E={{s-4M_W^2}\over{4M_W}}
\ee
\be
\beta_M={2\over{\sqrt{s}}}\left (\sqrt{{M_W\over 2}(\sqrt{E^2+\Gamma_W^2}+E)}+
i\sqrt{{M_W\over 2}(\sqrt{E^2+\Gamma_W^2}-E)}\right )
\ee
where $\sqrt{s_{1,2}}$ are the W boson invariant masses and $\sqrt{s}$ is
the center of mass energy. 
\par
The general first-order formula for the Coulomb correction then reads:
\be
F_{coul}={\alpha_{QED}\sqrt{s}\over{4p}}\left (\pi-2\delta^2 arctan\left 
({{s|\beta_M+\Delta |^2-4p^2}\over{4p\sqrt{s}Im(\beta_M)}}\right )\right )
\ee
\par
where for the two parameters $\Delta$ and $\delta$ we have included the 
following options:

\begin{itemize}

\item {\tt icoul}=1:  $\Delta =|s_1-s_2|/s$ and $\delta =1$ 

In this case  the above formula reproduces the one of Ref.~\cite{icoul1}
which was the only one 
available
in the previous version of \wphnosp.

\item {\tt icoul}=2:  $\Delta =|s_1-s_2|/s$ and $\delta =(1-2p/\sqrt{s})$

This factor $\delta$ implements the so called Khoze Chapovsky 
ansatz~\cite{icoul2} which
mimics the effect of non factorizable WW corrections. They are known to
be almost negligible in total cross sections but relevant for quantities
as W angle and mass distributions.

\item {\tt icoul}=3:  $\Delta =0$ and $\delta =1$ 

With this choice, the results of the Coulomb correction of 
Ref.~\cite{icoul1} agree with those obtained in Ref.~\cite{icoul3}, 
as therein explained. 

\item {\tt icoul}=4:  $\Delta =0$ and $\delta =(1-2p/\sqrt{s})$

This corresponds to the Khoze Chapovsky ansatz on the Coulomb correction of  
Ref.~\cite{icoul3}.

\end{itemize}
\par

The Coulomb factor $F_{coul}$, strictly related to the WW subset of 
diagrams (CC03), is implemented in the 
program as an additive correction 
\be\label{eq:coul}
|M_{CC,Mix}|^2=|M_{CC,Mix}^{Born}|^2+F_{coul}|M_{CC03}^{Born}|^2
\ee

\subsubsection{Fermion Loop}\label{fermiloop}

The inclusion of weak boson finite-width effects requires a careful treatment,
as these effects are strictly related to the gauge invariance of the theory
and even tiny violations of Ward identities can lead to totally
wrong predictions in many cases.

\begin{figure}[htb]
\begin{center}
\begin{picture}(130,120)(0,0)
\ArrowLine(10,100)(40,100)
\ArrowLine(40,100)(110,100)
\ArrowLine(110,20)(40,20)
\ArrowLine(40,20)(10,20)
\ArrowLine(115,40)(95,60)
\ArrowLine(95,60)(115,80)
\Photon(40,100)(50,75){2}{4}
\Photon(40, 20)(50,45){2}{4}
\ArrowLine(50,75)(50,45)
\ArrowLine(50,45)(75,60)
\ArrowLine(75,60)(50,75)
\Photon(75,60)(95,60){2}{3}
\Vertex(40,100){1.2}
\Vertex(40,20){1.2}
\Vertex(50,45){1.2}
\Vertex(50,75){1.2}
\Vertex(95,60){1.2}
\Vertex(75,60){1.2}
\put(08,100){\makebox(0,0)[r]{$e^-$}}
\put(08,20){\makebox(0,0)[r]{$e^+$}}
\put(112,100){\makebox(0,0)[l]{$e^-$}}
\put(112,20){\makebox(0,0)[l]{$\bar{\nu}_e$}}
\put(117,80){\makebox(0,0)[l]{$u$}}
\put(117,40){\makebox(0,0)[l]{$\bar{d}$}}
\put(41,85){\makebox(0,0)[r]{$\gamma$}}
\put(42,35){\makebox(0,0)[r]{$W$}}
\put(61,70){\makebox(0,0)[bl]{\shortstack{$d,s$\\$e,\mu,\tau$}}}
\put(57,51){\makebox(0,0)[tl]{\shortstack{$u,c$\\$\nu_e,\nu_\mu,\nu_\tau$}}}
\end{picture}
\qquad
\begin{picture}(130,120)(0,0)
\ArrowLine(10,100)(40,100)
\ArrowLine(40,100)(110,100)
\ArrowLine(110,20)(40,20)
\ArrowLine(40,20)(10,20)
\ArrowLine(115,40)(95,60)
\ArrowLine(95,60)(115,80)
\Photon(40,100)(50,75){2}{4}
\Photon(40, 20)(50,45){2}{4}
\ArrowLine(50,45)(50,75)
\ArrowLine(75,60)(50,45)
\ArrowLine(50,75)(75,60)
\Photon(75,60)(95,60){2}{3}
\Vertex(40,100){1.2}
\Vertex(40, 20){1.2}
\Vertex(50, 45){1.2}
\Vertex(50, 75){1.2}
\Vertex(95,60){1.2}
\Vertex( 75,60){1.2}
\put(08,100){\makebox(0,0)[r]{$e^-$}}
\put(08,20){\makebox(0,0)[r]{$e^+$}}
\put(112,100){\makebox(0,0)[l]{$e^-$}}
\put(112,20){\makebox(0,0)[l]{$\bar{\nu}_e$}}
\put(127,80){\makebox(0,0)[l]{$u$}}
\put(127,40){\makebox(0,0)[l]{$\bar{d}$}}
\put(41,85){\makebox(0,0)[r]{$\gamma$}}
\put(42,35){\makebox(0,0)[r]{$W$}}
\put(61,70){\makebox(0,0)[bl]{\shortstack{$u,c$\\$\nu_e,\nu_\mu,\nu_\tau$}}}
\put(57,48){\makebox(0,0)[tl]{\shortstack{$d,s$\\$e,\mu,\tau$}}}
\end{picture}
\end{center}
\vskip -1cm
\caption[]{The extra fermionic diagrams needed to cancel the 
terms which break gauge invariance.}
\label{extra_diagrams_eeenuud}
\end{figure}
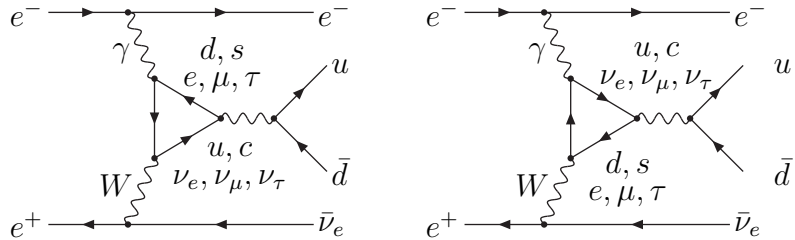

The most appealing approach which preserves gauge invariance is the Fermion
Loop (FL) scheme \cite{bhf} which
consists in the resummation of the fermionic one--loop corrections 
to the vector boson propagators and in the 
inclusion of all remaining fermionic one--loop corrections, in particular 
those to the Yang-Mills vertices (see \fig{extra_diagrams_eeenuud}).
Several realizations of this general scheme have been proposed and used in
numerical computations.

The so called Imaginary Fermion Loop (IFL), includes only the 
imaginary part of the loops, which is the minimal 
set of one--loop contributions needed to preserve gauge invariance.
The complete FL includes instead all contributions from 
fermionic loop corrections.

Initially applied only to the case where weak currents that couple to 
fermionic loops are conserved, both IFL and FL have been recently extended 
to non-conserved currents, which are present when one takes into account
external fermion  masses \cite{ifl,efl}.

\begin{figure}[t]
\vspace{0.2cm}
\bqas
\ba{ccc}
\vcenter{\hbox{
  \SetScale{0.7}
  \begin{picture}(110,100)(0,0)
  \ArrowLine(50,120)(0,140)
  \ArrowLine(100,140)(50,120)
  \ArrowLine(0,0)(50,20)
  \ArrowLine(50,20)(100,0)
  \ArrowLine(100,110)(80,70)
  \ArrowLine(80,70)(100,30)
  \Photon(50,20)(50,70){2}{7}
  \Photon(50,70)(50,120){2}{7}
  \Photon(50,70)(80,70){2}{7}
  \Text(-14,98)[lc]{$e^+$}
  \Text(77,98)[lc]{$\barnu_e$}
  \Text(-14,0)[lc]{$e^-$}
  \Text(77,0)[lc]{$e^-$}
  \Text(77,77)[lc]{$\bard$}
  \Text(77,21)[lc]{$u$}
  \Text(18,60)[lc]{$\wb$}
  \Text(11,26)[lc]{$\gamma,\zb$}
  \end{picture}}}
&\quad+&
\vcenter{\hbox{
  \SetScale{0.7}
  \begin{picture}(110,100)(0,0)
  \ArrowLine(50,120)(0,140)
  \ArrowLine(65,126)(50,120)
  \ArrowLine(100,140)(65,126)
  \ArrowLine(0,0)(50,20)
  \ArrowLine(50,20)(100,0)
  \ArrowLine(100,110)(80,70)
  \ArrowLine(80,70)(100,30)
  \Photon(50,20)(50,120){2}{7}
  \Photon(65,126)(80,70){2}{7}
  \Text(77,77)[lc]{$\bard$}
  \Text(77,21)[lc]{$u$}
  \Text(54,76)[lc]{$\wb$}
  \Text(11,66)[lc]{$\gamma,\zb$}
  \Text(-14,98)[lc]{$e^+$}
  \Text(77,98)[lc]{$\barnu_e$}
  \Text(-14,0)[lc]{$e^-$}
  \Text(77,0)[lc]{$e^-$}
  \end{picture}}}
\ea
\eqas
\bqas
\ba{ccc}
\vcenter{\hbox{
  \SetScale{0.7}
  \begin{picture}(110,100)(0,0)
  \ArrowLine(50,120)(0,140)
  \ArrowLine(100,140)(50,120)
  \ArrowLine(0,0)(50,20)
  \ArrowLine(50,20)(100,0)
  \ArrowLine(100,90)(50,90)
  \ArrowLine(50,90)(50,50)
  \ArrowLine(50,50)(100,50)
  \Photon(50,20)(50,50){2}{7}
  \Photon(50,90)(50,120){2}{7}
  \Text(77,72)[lc]{$\bard$}
  \Text(77,26)[lc]{$u$}
  \Text(22,52)[lc]{$u$}
  \Text(10,20)[lc]{$\gamma,\zb$}
  \Text(17,77)[lc]{$\wb$}
  \Text(-14,98)[lc]{$e^+$}
  \Text(77,98)[lc]{$\barnu_e$}
  \Text(-14,0)[lc]{$e^-$}
  \Text(77,0)[lc]{$e^-$}
  \end{picture}}}
&\quad+&
\vcenter{\hbox{
  \SetScale{0.7}
  \begin{picture}(110,100)(0,0)
  \ArrowLine(50,120)(0,140)
  \ArrowLine(100,140)(50,120)
  \ArrowLine(0,0)(50,20)
  \ArrowLine(50,20)(100,0)
  \ArrowLine(100,90)(50,50)
  \Line(50,90)(65,78)
  \ArrowLine(90,58)(100,50)
  \ArrowLine(50,50)(50,90)
  \Photon(50,20)(50,50){2}{7}
  \Photon(50,90)(50,120){2}{7}
  \Text(77,72)[lc]{$\bard$}
  \Text(77,21)[lc]{$u$}
  \Text(22,52)[lc]{$d$}
  \Text(10,20)[lc]{$\gamma,\zb$}
  \Text(17,77)[lc]{$\wb$}
  \Text(-14,98)[lc]{$e^+$}
  \Text(77,98)[lc]{$\barnu_e$}
  \Text(-14,0)[lc]{$e^-$}
  \Text(77,0)[lc]{$e^-$}
  \end{picture}}}
\ea
\eqas
\bqas
\ba{ccc}
\vcenter{\hbox{
  \SetScale{0.7}
  \begin{picture}(110,100)(0,0)
  \ArrowLine(50,120)(0,140)
  \ArrowLine(100,140)(50,120)
  \ArrowLine(0,0)(50,20)
  \ArrowLine(50,20)(100,0)
  \ArrowLine(100,110)(80,70)
  \ArrowLine(80,70)(100,30)
  \Photon(50,20)(50,120){2}{7}
  \Photon(18,132)(80,70){2}{7}
  \Text(77,77)[lc]{$\bard$}
  \Text(77,21)[lc]{$u$}
  \Text(48,71)[lc]{$\wb$}
  \Text(21,56)[lc]{$\zb$}
  \Text(-14,98)[lc]{$e^+$}
  \Text(77,98)[lc]{$\barnu_e$}
  \Text(-14,0)[lc]{$e^-$}
  \Text(77,0)[lc]{$e^-$}
  \end{picture}}}
&\quad+&
\vcenter{\hbox{
  \SetScale{0.7}
  \begin{picture}(110,100)(0,0)
  \ArrowLine(50,120)(0,140)
  \ArrowLine(100,140)(50,120)
  \ArrowLine(0,0)(50,20)
  \ArrowLine(50,20)(100,0)
  \ArrowLine(100,110)(80,70)
  \ArrowLine(80,70)(100,30)
  \Photon(50,20)(50,120){2}{7}
  \Photon(64,15)(80,70){2}{7}
  \Text(77,77)[lc]{$\bard$}
  \Text(77,21)[lc]{$u$}
  \Text(40,41)[lc]{$\wb$}
  \Text(19,56)[lc]{$\wb$}
  \Text(-14,98)[lc]{$e^+$}
  \Text(77,98)[lc]{$\barnu_e$}
  \Text(-14,0)[lc]{$e^-$}
  \Text(77,0)[lc]{$e^-$}
  \end{picture}}}
\ea
\eqas
\vspace{-2mm}
\caption[]{The $t$-channel component of the CC20 family of diagrams: fusion,
bremsstrahlung and multi-peripheral.}
\label{cc20gz}
\end{figure}
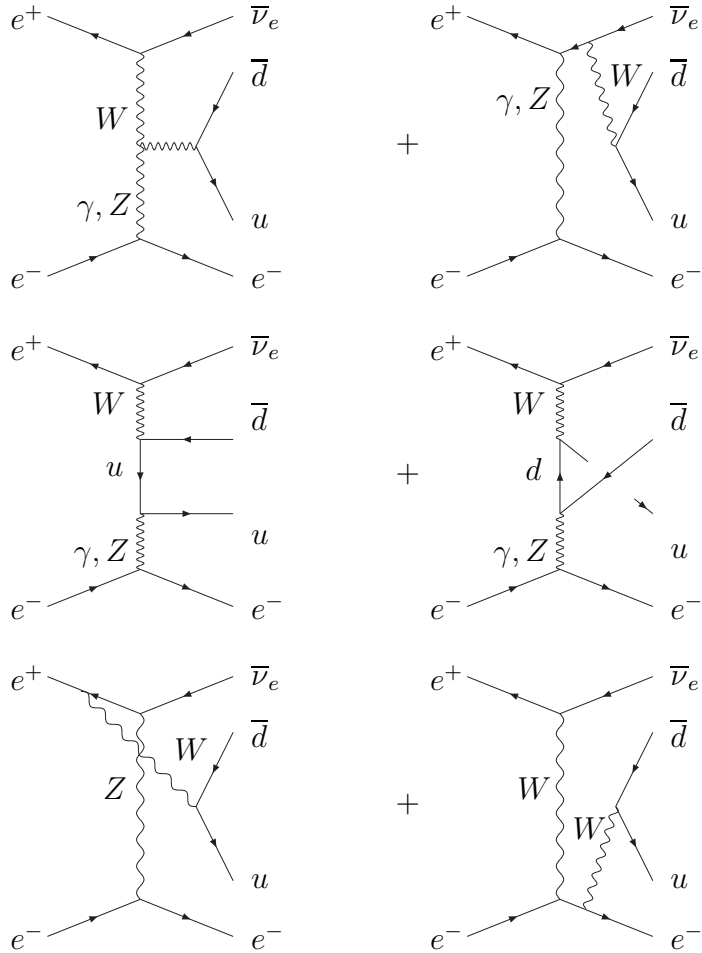

A class of processes where these corrections become particularly relevant
is the one related to single W production, for example 
$e^+e^-\rightarrow e^-\bar\nu_e u\bar d$ whose complete set of $t$-channel 
diagrams is shown in \fig{cc20gz}. These processes, dominated by the four
$t$-channel photon diagrams when the electron is emitted at very small angle,
diverge in the massless limit $m_e\rightarrow 0$. Moreover, their apparent
$q_\gamma^{-4}$ behaviour gets reduced to $q_\gamma^{-2}$ by gauge 
cancellations. If the electron mass is taken into account there are however
terms proportional to $m_e^2/q_\gamma^{4}$. 
This kind of processes therefore requires the inclusion of
fermion masses and the use of a gauge preserving scheme when taking into
account finite width effects.

The IFL scheme for non-conserved currents has been implemented, as an option,
in the new version of \wphnosp.

As an alternative option, we have 
also included in \wph a simplified gauge-preserving scheme.
An easier way to preserve gauge invariance, commonly adopted for practical 
purposes, is in fact the use of the so called 
{\tt{fixed width}} approximation (FW), which gives unphysical widths for 
space-like momenta but retains U(1) gauge invariance. 
In the unitary gauge, this {\it{naive}} scheme consists in
replacing $M_W^2$ with $M_W^2-iM_W\Gamma_W$ both in the denominator and in the
$p^\mu p^\nu$ term of the bare W propagator. Note that neglecting to modify 
the latter could lead to large errors which increase with energy. This 
effect, negligible at LEP2, could be greatly enhanced in the high energy 
range of future colliders. 

A comparison between the IFL and FW schemes for single W production at small 
angle (collinear electron), where one expects gauge invariance issues to be 
essential, has shown very good  agreement for total cross sections
but not for distributions.

\subsubsection{Running of $\alpha_{QED}$}\label{r_alpha}

The running of the electromagnetic coupling constant $\alpha_{QED}$ can give 
rise to important corrections, especially for small momentum transfer 
($q^2\simeq 0$) kinematical configurations as in small angle scattering and
small invariant mass regions.  
\par
The $G_\mu$ scheme, which fixes the value of $\alpha_{QED}$ to 
be $\alpha_{G_F}\simeq 1/132$, is appropriate for high energy 
$s$-channel scattering processes,
like WW, ZZ and $h$Z vector boson pair production. 
However, it is not adequate for multi--scale processes as
single W and single Z production which are determined by both the vector 
boson mass scale and the small transfered momentum  $|q_\gamma^2|$ in the 
$ee\gamma$ vertex.

As it is known, the choice of the correct scale for the couplings is related
to the real part of Fermion Loop radiative corrections.
Presently, the complete FL has been computed only for single W processes 
\cite{efl}, leading to a drastic reduction of the theoretical uncertainties 
on the cross section, and is implemented exclusively in {\tt WTO}\cite{wto}. 
Since a detailed analysis of these corrections has shown that the running 
of $\alpha_{QED}$ accounts for a large  part of FL effects\cite{4f},
we have included in the new version of \wph two additional options for the 
determination of $\alpha_{QED}$. 

For CC processes with an electron (or positron) in the final state, a 
well-behaved, even if {\it ad} {\it hoc}, prescription for the running of 
$\alpha_{QED}$ can be selected via the input parameter {\tt ialftsw} 
(see Sec.~\ref{corrections}).
Since $s$- and $t$-channel diagrams constitute two independent gauge invariant 
subsets, it is possible to use two different values of $\alpha_{QED}$: 
$\alpha_{QED}(q_\gamma^2)$ for the $t$-channel (applied to two vertices), 
and $\alpha_{G_F}$ for the $s$-channel. 
In this way, $SU(2)\times U(1)$ is preserved.
A comparison between the results obtained with FL and IFL plus running 
$\alpha_{QED}$ has shown that the agreement between the two schemes, even if 
process dependent, is generally better 
than 2-3$\%$ for single W production at LEP2,
i.e. within the present experimental precision tag.

More generally, it has been also introduced the possibility of having
$\alpha_{QED}$-running in all photonic vertices $\gamma f\bar f$ 
(see also Sec.~\ref{corrections}). For each vertex the scale is taken equal to
the square of the photon momentum.
This approach does not
preserve gauge invariance. However, at LEP2 energies, it can give an 
adequate description of NC and Mixed processes dominated by low $q^2$ 
photons, such as single Z, Z$\gamma^*$ or $\gamma^*\gamma^*$. 
An example of the relevance of such an approach can be found
in fig.~\ref{singleZq}.  In the first two plots to the right, cross sections
with $\alpha _{em}$ running are compared with the analogous ones in $G_\mu$
scheme for single Z processes, which are dominated
by low scale photon propagator contributions.
In such a case one can find differences between the two approaches of the
order of 5\%.

\subsection{ Resonances and interface to hadronization}

In \wphtwo there is the possibility to account for the resonant structures
in $\gamma^* \ar q\bar q$ near  threshold through a link
to {\tt R\_RES}\cite{boonek}. These routines provide a parametrization
of the low mass $q\bar q$ resonances and continuum  which reproduces the 
R-factor. This feature is useful for phase space regions where the 
$\gamma^* \gamma^*$ or Z$\gamma^*$ contributions are dominant. However,
in final states with an $e^+e^-$ pair, low $m(q\bar q)$ dominant contributions
may also come from   multiperipheral diagrams when at least one electron
is produced at small angle.  In this case the above parametrization may 
not be adequate. One may however decide
to evaluate the two contributions, applying  
appropriate cuts, in two different runs, and eventually generate  
events simultaneously in the two regions with the  one-shot procedure
described in section~\ref{oneshot}.

Hadronization in \wph is provided via a link to \pytnosp . The interface
is provided by the routine {\tt AB\_LU4FRM} which is derived from
 \pytnosp 's {\tt PY4FRM}.  
When the final state consists of two identical pairs of quarks 
or of four quarks in a Mixed 
final state, different color singlets can be formed. In the first case, 
\wphtwo evaluates for every generated phase space point the separate
contributions of the diagrams corresponding to the different pairing,
and assigns  the color flow  with a probability proportional to their
weights. In the Mixed case, the separate generation  of charged
and neutral currents, which is the standard procedure in \wphnosp ,  
is used to get the correct proportion of colour pairings. 

The hadronization of quark pairs with $m(q\bar q) < 2$ GeV is not handled 
by \pytnosp, but by the routine {\tt Hadgen} of the {\tt R\_RES} package.
The description of this region by theoretical models is in fact known
to be problematic, while {\tt Hadgen} produces final states according
to measured exclusive $e^+e^-$ cross sections compiled for this purpose.

\subsection{Susy and Anomalous couplings}\label{susy}
Besides SM Higgs processes with two $b$'s ( $c$'s or $\tau$'s ) plus two 
other fermions in the final state, \wph computes also SUSY neutral Higgs 
production in the same final channels. 
\par
As for other previously discussed processes, in the new version of \wph the 
contributions of all diagrams containing the Higgs can be optionally 
separated from the rest or individually selected, as explained in more detail 
in Sec.~\ref{higgs}.
\par
Moreover, in \wphtwo we have introduced new options for the SUSY parameters 
to be given as input (see also Sec.~\ref{higgs}). We have in fact updated the 
one--loop corrections to the lightest neutral Higgs mass $M_h$ used 
in the previous version of the program, by including the two-loop RG improved effective 
potential results as given in Ref.~\cite{2loophiggs}.
As a further option, the user can also give directly the three parameters
$M_h$, $M_A$ and $tg(\beta)$ as independent inputs, where $M_A$ is the 
CP-odd Higgs mass and $tg(\beta)$ is the ratio between the two vacuum 
expectation values.

As far as Anomalous Couplings effects are concerned, we have implemented
those relative to the trilinear vertices WW$\gamma$ and WWZ. 
Among the various possible parametrizations, we have included the one given 
in Ref.~\cite{ancoup}. 
\par
In \wphtwonosp, we have extended the implementation of Anomalous Couplings to 
the new fully massive matrix elements.

\subsection{Distributions and unweighted event generation}\label{unweigh}

Even if no particular new features have been introduced as far as
distributions and unweighted event generation of a single process are concerned,
we recall here their main properties for completeness and as
an introduction to the next subsection. 

 \wph has a built-in mechanism for computing any distribution at parton
level.
To this aim, the user has to  define the variables for which differential
cross sections need to be produced in the routine {\tt fxn.f},
and to specify in the \incas\ their number, their sub--intervals and binnings.
The program produces them
automatically during integration. 

The results for all distributions, together with 
an estimate of the error
for each bin, are written in the file {\tt ABDIS.DAT}. 

Distributions can also be produced generating 
unweighted events. This is particularly useful in experimental studies
involving hadronization and detector simulations. 
\wph can produce unweighted events both at parton and at 
hadron level (with a link to \pytnosp). 
In the new 2.0 version, it can also produce 
unweighted events for as many simultaneous massive processes as desired. 
This last feature is described in the next section. 
    
As explained in \cite{wph1}, unweighted events for a single process at a time
can be produced with the hit-or-miss method while 
evaluating the integral with \vegnosp. In this case the last but one  effective
iteration after thermalization is used to find the maximum {\tt rmaxfxn} of the 
integrand and the last one to generate unweighted events.\footnote 
{It has to be
noticed in this respect that in version 1.0 the search for the maximum
was performed in the first effective iteration and the generation in
the second one, so that the user was forced to use
only two iterations when producing unweighted events.
We have removed this restriction as we find sometimes useful
to have some more iterations (other than thermalization)
before actually generating events in order to fully exploit \veg adaptivity.}
The drawbacks of this procedure are that the number of generated events
cannot be predicted and that one may have some generated events whose weight
is larger than {\tt rmaxfxn}$\times${\tt scalemax} (where {\tt scalemax} is the
factor used to tune the efficiency of the hit-or-miss selection).
We therefore recommend the use of a two step procedure for generation:
after a first run with {\tt iflat=1},  {\tt istorvegas=1} and {\tt irepeat=0}
one can perform a second run with {\tt iflat=1} and {\tt irepeat=2}
(cfr. Section~\ref{input})
in which one requires, via {\tt nflevts}, 
a predefined number of events. These will be produced using the 
maximum and the integration grid determined in the 
last iteration of the first run. 

Usually in such a two step procedure, one does not require 
hadronization and does not store the events generated in the first run,
which then simply evaluates the integral and determines
the ``best'' adaptive grid. The necessary information for actual generation 
is stored in the
file {\tt ABVEGAS.DAT}. It is obvious that any  such file can be 
reused for generation as many times as one wants. The multi--process
simultaneous generation described in the next section is in practice just
a way of combining and using the information of many such files
at the same time. It can also be used for only one process and in such
a case it constitutes a convenient replacement for the second run above.

\subsection{One--shot generation}\label{oneshot}
Besides the event generation described in the previous section and already 
present in the first version of \wphnosp, a much more powerful method has been 
implemented in \wphtwo for all completely massive processes.
  This allows the user to generate unweighted events
not on a process by process basis but for any possible set of
processes and channels and cuts in a single run, 
giving eventually a complete event sample in which
all included 4-fermion final states are produced with the correct 
frequency.

In order to use this feature, single run inputs (hereafter we will call them
\infis) must be present. The best way to understand what they are
is just to  describe how they can be produced. 
For every process (and cuts) one wants to 
consider, one has to perform a run exactly as the first 
of the two step procedure  described in the previous section. 
This is by all practical
means a normal integration run with the flags 
{\tt iflat=1, istorvegas=1 irepeat=0} (cfr. Section~\ref{input}). 
One then appends the normal output file with integration results of every run
to the corresponding {\tt ABVEGAS.DAT} and saves it with an appropriate name
(say {\tt proc1.inp, proc2.inp,...}).
Each such \infi\ will therefore contain all information about the input 
of the first run (process, energy, options, cuts, etc.), as well as the result 
of the numerical integration, the best grid and the maximum 
for the single process. 

If one now chooses to run in the ``one--shot'' mode 
({\tt ionesh=1}), one specifies in the input the number of the 
\infis\ and their names.
\wph will then read from the files all necessary information and build up
for every single process its probability and a new maximum normalized to it.
According to this probability, one final state will be extracted at a time.
For the selected process the relative cuts and options will be used,
and an event will be generated with a frequency in phase space determined
by its flags and grid, and this event will be compared with
the  normalized maximum in order to keep or reject it. 
The procedure
will then be repeated starting from the process selection until
the required number ({\tt nunwevts}) of unweighted events is produced.

The efficiency of this method relies heavily on an accurate preparation
of the \infis . The choice of the grids and of the maxima affects
the efficiency but not the correctness of the result, provided that one
eventually checks the number of events exceeding the maximum and
if necessary modifies {\tt scalemax}.    
 
As a consequence, one may use the same \infis\ for energies which are
different from the energy at which they were produced.
However, if the energy difference becomes too large the efficiency may
become unacceptably small. In the same spirit, one does not need to
produce a single \infi\ for every final state, and there is indeed the 
possibility to use only one representative process for every set of 
processes with the same {\tt iproc} value, whose matrix elements differ only 
for charge conjugation or for final particle masses. In this case the
representative process must be the one with {\tt ich=1}.
In our experience this faster possibility is quite safe and efficient.   

Any subset of the \infis\ can of course be used, so that once 
the files for all processes have been prepared, 
one may use them for partial generations as well. 
As an example one may imagine to generate only neutral current final states, 
or only processes involving electrons, and so on. Any final state may also be 
divided, with respect to \infis, in different exclusive complementary 
parts with different 
cuts. For instance one may think of separating regions with electrons lost in 
the beampipe from ZZ or WW contributions. Therefore one
can have several \infis\ 
for a single process and use them all together or not.

\section{Input} \label{input}
\vsk

\wphtwo is much more flexible in accepting input parameters than the
previous version of the program.
The syntax is almost identical to the one required by the
CERN library routine {\tt FFREAD}. New routines internal to \wph are however 
used ({\tt iread, rread}),
so that real variables can (and must) be given in double precision.
The difference between single and double precision can be relevant for
some cuts or parameters in particular regions.

All lines in the file of input (hereafter called \inca ) 
must not exceed 80 characters.
A {\tt *} or {\tt C} character at the beginning of a line identifies it
as a comment line. Comment lines can be freely interspersed within the \inca,
with the only obvious exception that they must not interrupt a list
of input values for a single array variable.
The name of the variable to be read must be specified as
the first word of a line 
(needs not to begin in column 1). Its value (values) must follow it. 
The list of values can span several lines.
Variables which are  not needed for the process under study 
will be ignored. They can be left in the \inca\ without harm.
All variables actually read from the \inca\ will
be reproduced in the output.
Since several related runs are often needed,
we find it convenient to have a master \inca.
For each run only a small number of variables
are modified and all other input values, including those which are
irrelevant for the run at hand, are left in place for possible further use.
The order in which variables appear in the \inca\ is immaterial.
The user may freely change it. 
In output, however, variables will be ordered as they are read by 
the program.
 
Every parameter  whose initial is {\tt i} or {\tt n} is of type
{\tt integer}. All  others are {\tt real*8}.  
When a variable has a yes/no option the value {\tt 1} corresponds to {\tt YES},
 {\tt 0} to {\tt NO}.
All energies and masses must be expressed in GeV.

We recall that \wphtwo can be used in two different modes. The program
can compute or generate events for a single process or it can generate events
for a user--selected list of processes, provided the corresponding \infis\
are available.
Different sets of inputs must be specified for the two modes.
In the following we discuss each input variable, its meaning and purpose
separating the single process and the one--shot cases.
\par

We begin obviously with:
\vsk
{\large \tt ionesh}: this yes/no flag determines whether the program is supposed
(1) to generate unweighted events for several massive processes simultaneously, 
using a list of \infis, or it is required (0) to compute only one 
(massless or massive) process.
\Infis\ are prepared using this second option.

\vsk
\subsection{One process case: {\tt ionesh=0}} \label{ionesh0}
\vsk
If {\tt ionesh=0} the following flags are required:
\vsk
{\large \tt e\_cm}: the center of mass energy.
\vsk
{\large\tt imass}: if {\tt iproc} $\leq$ 32
{\tt imass=0} means massless fermions and {\tt imass=1} means fully massive
fermions.
If {\tt iproc} $\geq$ 33, which 
corresponds to the Higgs signal and its irreducible background processes,
{\tt imass} is ineffective: 
the final states of Table 3 have $b$ massive in processes 33-39, 
$c$ massive in processes 40-46 and $\tau$ massive in processes 47-53 while
all other external particles are massless.

\subsubsection{Selection of the final state} \label{finalstate}
\vsk
{\large\tt iproc,ich}:   
the possible values for  these two parameters can be found in  Tables 1,2,3.
{\tt iproc}  selects a group of processes with the same Feynman diagrams,
while {\tt ich} distinguishes
a specific final state within the group. The 
choice of {\tt ich} is relevant also in the massless case, if the program is 
used as an event generator and hadronization is performed via \pyt
or cuts which differentiate between different particle types are used. 
\vsk
{\large\tt ickm}: 
this flag determines whether the 
Cabibbo--Kobayashi--Maskawa matrix is taken into account ({\tt ickm=1}) or
it is taken equal to the unit matrix ({\tt ickm=0}). As far as the three
processes $ e^+ e^- \ar   s  \bar u  u  \bar s,\  
  d  \bar c  c  \bar d,\ b  \bar c  c  \bar b $
are concerned, the CC exchange contribution is obtained with {\tt ickm=1} 
and {\tt iproc=5}, 
     {\tt ich=21,22,23}, while the usual NC part is obtained with 
     {\tt iproc=24} and {\tt ich=1,2,4} respectively. 
     The interference between these two contributions is ignored. 

\subsubsection{Selection of a subset of Feynman diagrams} \label{subset}
\vsk
In some cases the user may want to compute only a subset of the Feynman diagrams
which describe the production of a physical final state. This is useful for
instance when heavy intermediate particles can go on their mass--shell.
The most commonly used subsets can be obtained with the flags described below.
One should however be aware that many of the subsets are not gauge invariant.
The results should therefore be taken with great care, bearing in mind that
our matrix elements are computed in the Unitary Gauge.
\vsk
{\large\tt iccnc}: in mixed processes this flag allows the user to compute
the  CC ({\tt iccnc=1}), NC ({\tt iccnc=2}) or full CC+NC+interference
({\tt iccnc=3}) contribution. Let us recall that, if {\tt iccnc=3}, 
\wph by default integrates separately the two contributions, CC and NC, 
adding the interference to one of them. 
When generating events ({\tt iflat=1}), in order to avoid negative weights
we adopt the following procedure, which regards as well {\tt iccnc=1} and 
{\tt iccnc=2}. The interference is initially summed to the 
CC contribution (which is generally the larger one). If the result is  
negative, the quantity CC+Interference is set to zero while the full 
CC+NC+Interference contribution is assigned to the NC part.
For one--shot generation it is more efficient to have separate 
\infis\ for the 
CC ({\tt iccnc=1}) and NC ({\tt iccnc=2}) parts.  For this reason
only this possibility has been implemented and must be used.

\begin{figure}[t]
{
\unitlength=1.0 pt
\SetScale{1.0}
\SetWidth{0.7}      
\normalsize    
\noindent
{}\hspace{-20pt}
\begin{picture}(190,198)(0,0)
\Text(30.0,160.0)[r]{$e$}
\ArrowLine(32.0,160.0)(74.0,160.0) 
\Text(94.0,166.0)[b]{$a$}
\DashLine(74.0,160.0)(116.0,160.0){3.0} 
\Text(166.0,180.0)[l]{$p3$}
\ArrowLine(116.0,160.0)(158.0,180.0) 
\Text(166.0,140.0)[l]{$p4$}
\ArrowLine(158.0,140.0)(116.0,160.0) 
\Text(66.0,120.0)[r]{$e$}
\ArrowLine(74.0,160.0)(74.0,80.0) 
\Text(30.0,80.0)[r]{$\bar{e}$}
\ArrowLine(74.0,80.0)(32.0,80.0) 
\Text(94.0,86.0)[b]{$b$}
\DashLine(74.0,80.0)(116.0,80.0){3.0} 
\Text(166.0,100.0)[l]{$p5$}
\ArrowLine(116.0,80.0)(158.0,100.0) 
\Text(166.0,60.0)[l]{$p6$}
\ArrowLine(158.0,60.0)(116.0,80.0) 
\end{picture} \
\hfil 
\begin{picture}(190,198)(0,0)
\Text(30.0,160.0)[r]{$e$}
\ArrowLine(32.0,160.0)(74.0,160.0) 
\Text(94.0,166.0)[b]{$b$}
\DashLine(74.0,160.0)(116.0,160.0){3.0} 
\Text(166.0,180.0)[l]{$p5$}
\ArrowLine(116.0,160.0)(158.0,180.0) 
\Text(166.0,140.0)[l]{$p6$}
\ArrowLine(158.0,140.0)(116.0,160.0) 
\Text(66.0,120.0)[r]{$e$}
\ArrowLine(74.0,160.0)(74.0,80.0) 
\Text(30.0,80.0)[r]{$\bar{e}$}
\ArrowLine(74.0,80.0)(32.0,80.0) 
\Text(94.0,86.0)[b]{$a$}
\DashLine(74.0,80.0)(116.0,80.0){3.0} 
\Text(166.0,100.0)[l]{$p3$}
\ArrowLine(116.0,80.0)(158.0,100.0) 
\Text(166.0,60.0)[l]{$p4$}
\ArrowLine(158.0,60.0)(116.0,80.0) 
\end{picture} \ 
}
\vspace{-13mm}
\caption[]{The sets of diagrams which can be singled out in NC
           processes with the flags {\tt izz}, {\tt inc08}, {\tt izg},
           {\tt izg34}, {\tt izg56}, {\tt igamgams} are generated by the two 
            diagrams above for different choices of
           the vector bosons $a$ and $b$. If identical particles are present
           in the final state there are two additional diagrams which can be obtained
           exchanging $p3$ with $p5$. In this case the number of diagrams selected
           is twice the number mentioned in \ref{subset}. }
\label{nc02}
\end{figure}
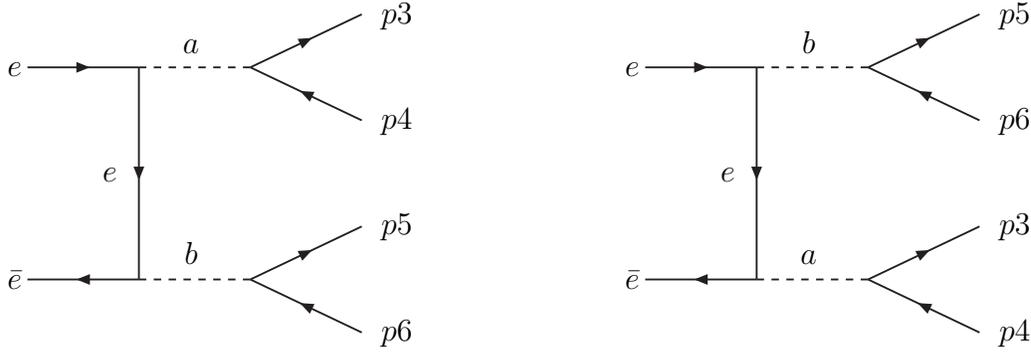

\vsk
\par
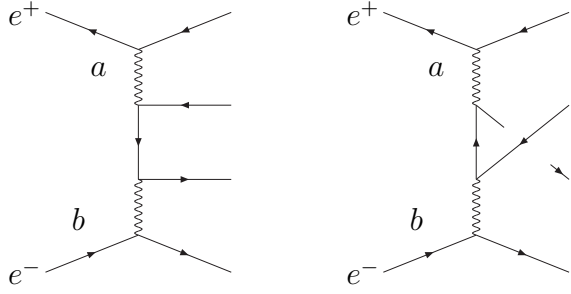
\begin{figure}[bt]
\vspace{0.2cm}
\bqas
\ba{cc}
\vcenter{\hbox{
  \SetScale{0.7}
  \begin{picture}(110,100)(0,0)
  \ArrowLine(50,120)(0,140)
  \ArrowLine(100,140)(50,120)
  \ArrowLine(0,0)(50,20)
  \ArrowLine(50,20)(100,0)
  \ArrowLine(100,90)(50,90)
  \ArrowLine(50,90)(50,50)
  \ArrowLine(50,50)(100,50)
  \Photon(50,20)(50,50){2}{7}
  \Photon(50,90)(50,120){2}{7}
  \Text(10,20)[lc]{$b$}
  \Text(17,77)[lc]{$a$}
  \Text(-14,98)[lc]{$e^+$}
  \Text(-14,0)[lc]{$e^-$}
  \end{picture}}}
&
\vcenter{\hbox{
  \SetScale{0.7}
  \begin{picture}(110,100)(0,0)
  \ArrowLine(50,120)(0,140)
  \ArrowLine(100,140)(50,120)
  \ArrowLine(0,0)(50,20)
  \ArrowLine(50,20)(100,0)
  \ArrowLine(100,90)(50,50)
  \Line(50,90)(65,78)
  \ArrowLine(90,58)(100,50)
  \ArrowLine(50,50)(50,90)
  \Photon(50,20)(50,50){2}{7}
  \Photon(50,90)(50,120){2}{7}
  \Text(10,20)[lc]{$b$}
  \Text(17,77)[lc]{$a$}
  \Text(-14,98)[lc]{$e^+$}
  \Text(-14,0)[lc]{$e^-$}
  \end{picture}}}
\ea
\eqas
\vspace{-2mm}
\caption[]{The multi-peripheral type diagrams.}
\label{multiper}
\end{figure}

For the following group of 
flags there are four possible values. If $F$ is the full set
of Feynman diagrams for the process and  $S$ is the relevant subset
\begin{itemize}
\item {\it flag} {\tt=0} means computing $\vert F\vert^2$
\item {\it flag} {\tt=1} means computing $\vert S\vert^2$
\item {\it flag} {\tt=-1} means computing $\vert (F-S)\vert^2$
\item {\it flag} {\tt=2} means computing $\vert F\vert^2 - \vert S\vert^2$
\end{itemize}
These four possible values and the corresponding subsets are implemented
only for massive matrix elements ({\tt imass=1}). In the massless
case ({\tt imass=0}), one must set {\it flag} {\tt=0},
with the only exception of {\tt icc3} where {\tt icc3}=1 can
also be selected. 
\vsk
{\large\tt icc3}: the subset is CC03. 
\vsk
{\large\tt izz}: the subset is NC02 (see \fig{nc02} with $a=b=$Z). 
\vsk
{\large\tt inc08}: the subset is NC08 (see \fig{nc02} with $a,b=$Z,$\gamma$).
\vsk
{\large\tt izg}: the subset is the set of four NC08 diagrams in which one pair
                 of final state fermions is connected to an 
intermediate Z--boson
                 and the other to an intermediate photon
                 (see \fig{nc02} with $a=$Z,$b=\gamma$ or $a=\gamma,b=$Z)
\vsk
{\large\tt izg34}: the subset is the set of two NC08 diagrams in which the pair
                 of final state fermions with momenta {\tt p3} and {\tt p4} 
is connected
                 to an intermediate  photon
                 and the other to an intermediate Z--boson.
                 (see \fig{nc02} with $a=\gamma,b=$Z)
\vsk
{\large\tt izg56}: the subset is the set of two NC08 diagrams in which the pair
                 of final state fermions with momenta {\tt p5} and {\tt p6} is 
connected
                 to an intermediate  photon
                 and the other to an intermediate Z--boson
                 (see \fig{nc02} with $a=$Z,$b=\gamma$).
\vsk
{\large\tt igamgams}: the subset is NC02 (see \fig{nc02} with $a=b=\gamma$). 
\vsk
{\large\tt igamgam}: the subset is the set of multiperipheral diagrams 
characterized by two $t$-channel photon propagators (i.e. $\gamma\gamma$ channel,
see \fig{multiper} with $a=b=\gamma$).
\vsk
{\large\tt imulper}: the subset is the set of all multiperipheral diagrams characterized by two
                     $t$-channel neutral vector boson propagators 
                     (see \fig{multiper}
                     with all possible intermediate neutral vector bosons).
                     This flag is active only for NC processes or the NC part
                     of Mixed processes.  
\vsk
{\large\tt itch}: the subset is the set of so--called $t$-channel diagrams,
                 those in which there is
                 at least one $t$-channel vector boson propagator
                 (see for example \fig{cc20gz}).
\vsk
{\large\tt itchnomp}: the subset is the set of diagrams in which there is
                 at least one $t$-channel vector boson propagator but which 
                 are not of the multiperipheral type
                 (see \fig{multiper} and \fig{cc20gz}).
                     This flag is active only for NC processes or the NC part
                     of Mixed processes.  
\vsk

\subsubsection{Selection of a phase space mapping} \label{phasespace}
\vsk
{\large\tt ips\_cc, ips\_nc}:
these flags select among the various phase space
mappings for the integration over invariant masses.
{\tt ips\_cc} refers to the phase space
of CC or Higgs signal contributions. {\tt ips\_nc} to the NC 
contributions. Only one of the two flags is relevant if the process
is not of the Mixed (or NC+Higgs) type. For Mixed processes the two flags can be
chosen independently.
Both  parameters  can assume 4 values: {\tt 1} for double resonant mapping, 
{\tt 2} for  resonant mapping on the $34$ 
($36$ for CC part of Mixed processes) invariant mass and
non-resonant distribution of the $56$ ($54$ for CC part of Mixed) 
invariant mass,
{\tt 3} for  resonant mapping on the $56$ ($54$ for CC part of Mixed) 
 invariant mass and
non-resonant distribution of the $34$ ($36$ for CC part of Mixed)  
invariant mass,
{\tt 4} for non-resonant distribution of both invariant masses.
We recall that a resonant mapping is intended to flatten out the 
relevant mass peaks of the pair of final particles at hand.
For neutral currents this includes the Z resonance, together with a possible
photon enhancement at low invariant masses or an additional propagator 
enhancement corresponding to a Higgs resonance. 
\vsk
{\large\tt ismallangle}: it must be set {\tt 0} if no electrons or positrons
are present in the final state. 
{\tt ismallangle}$\neq${\tt 0}  selects the mapping of the 
angular distribution of 
final state electrons and/or positrons which can be strongly affected 
by $t$-channel enhancements. For such a reason we recommend to use
these latter values only together with {\tt imass=1}. 
\par
{\tt ismallangle=0} is appropriate when the final state $e^\pm$ are 
at large angles with respect to the beam.
\par
 {\tt ismallangle=1} is appropriate when
the process is dominated by final states with one $e^-$ (or $e^+$) close to its 
initial direction. If {\tt ismallangle=1} and there is at least one $e^-e^+$
 pair in the final state, the next flag {\tt ismallangle\_ee}
must be set.  
\par
{\tt ismallangle=2} is appropriate when 
the process contains at least one $e^-e^+$ pair in the final state and it
is dominated by final states with both $e^-$ and $e^+$ close to 
their initial direction.
\par
If {\tt ismallangle} is non zero {\tt ips\_nc}  ({\tt ips\_cc}) 
must be equal to {\tt 3} or {\tt 4} for NC or NC part of a Mixed process 
(CC or CC part of a Mixed process).  
\vsk
{\large\tt ismallangle\_ee}: only relevant if {\tt ismallangle=1} and 
{\tt iproc=7, 18, 21, 26, 28, 30}. 
{\tt ismallangle\_ee=3(4)} selects a phase space which is appropriate
for a final state $e^-(e^+)$ close to its initial direction.
According to our convention, before ISR the $e^+$ momentum is in the 
$+z$ direction. 
It has to be noticed that for the CC part of the Mixed process 
$e^-e^+\nu_e\bar\nu_e$, the choice {\tt ips\_cc=3} corresponds to
the W-resonant mapping on the $e^+\nu_e$ pair for {\tt ismallangle\_ee=3} 
(electron at small angle), and on $e^-\bar\nu_e$ for 
{\tt ismallangle\_ee=4} (positron at small angle).
\vsk
\subsubsection{Selection of electromagnetic radiation} \label{radiation}
\vsk
{\large\tt isr}: 
{\tt isr=0} means no Initial State Radiation (ISR). 
{\tt isr=1} means ISR using Structure Functions~\cite{isr1}. 
{\tt isr=2} means ISR using \QEDPS~\cite{qedps}.
\vsk 
{\large\tt ibeam}: 
yes/no beamstrahlung correction~\cite{circe}. This effect is relevant
only for Linear Collider energies. {\tt ibeam=0} must be chosen in
all other cases.
\vsk
\subsubsection{Higher order corrections and scheme selection} \label{corrections}
\vsk
{\large\tt icoul}: if {\tt icoul}=0, no Coulomb corrections to the 
CC03 part(\eqn{eq:coul}) are included. If one wants to compute them,
one can choose among 4 possible values (1,2,3,4). For their meaning one
has to refer to Sect~\ref{coulomb}.
\vsk
{\large\tt istrcor}:  yes/no 'naive' QCD corrections to the cross sections.
It has to be noticed
that if {\tt istrcor=0}, quarks are present in the final state  
and {\tt igwcomp,  igzcomp} or {\tt ighcomp} are equal to {\tt 1}
the widths are computed for consistency without QCD corrections.
\vsk
{\large\tt ifloop}: yes/no IFL corrections to CC processes or to the CC part
of Mixed processes with at least an electron (or positron) in the final state.
{\tt ifloop = 1} requires the W width to be $s$--dependent 
(see below {\tt ipr=2}).
{\tt ifloop = 1} is relevant for low angle electron final states and it
is therefore only implemented for the massive case {\tt (imass=1)}. 
\vsk
{\large\tt ialftsw}: yes/no running $\alpha_{QED}(q^2_\gamma)$ applied to two vertices in $t$-channel   
diagrams of CC processes or CC part of Mixed processes, with at least an electron (or positron) in the final state. Implemented only for the massive case 
{\tt (imass=1)}. 
\vsk
{\large\tt ialfar}: yes/no running $\alpha_{QED}(q^2_\gamma)$ applied to
all photon vertices $\gamma f\bar f$. This option, even if not gauge invariant,
can be
useful for processes dominated by electromagnetic contributions. Implemented only for the massive case {\tt (imass=1)}. 
\vsk
{\large\tt ipr}: this flag selects among running or constant Z, W, 
Higgs widths in $s$-channel propagators:\par
          {\tt ipr=0}  Z, W, H have constant width\par
          {\tt ipr=1}  Z, W, H have $s$--dependent width\par
          {\tt ipr=2}  Z, H have constant width; W has $s$--dependent width.\parno
If {\tt ifloop=1}, {\tt ipr} is set automatically to {\tt 2}. 
A warning will appear in the output.
\vsk
{\large\tt iswgcomp}: If this flag is set to {\tt 1}, $sin^2\theta_W$ and $g$
  are computed in terms of the Z mass, the W mass and  $G_f$.
  This corresponds to use the ``$G_\mu$ scheme''.
   If it is set to {\tt 0}, the values for $sin^2\theta_W$ and 
   $\alpha_{QED}$ are taken from the {\tt DATA} and
   $g^2=4 \pi \alpha_{QED} / sin^2 \theta_W$.
\vsk
{\large\tt igwcomp}: if {\tt igwcomp=1} the W width is computed using standard
formulae. If {\tt igwcomp=0 } the W width is taken from the {\tt DATA}.
\vsk
{\large\tt igzcomp}: if {\tt igzcomp=1} the Z width is computed using standard
formulae. If {\tt igzcomp=0 } the Z width is taken from the {\tt DATA}.
\vsk
{\large\tt ighcomp}: if {\tt ighcomp=1} the H width is computed using standard
formulae. If {\tt ighcomp=0 } the H width is taken from the {\tt DATA}.
\vsk
{\large\tt iresonance}: yes/no interface to 
 the routines of {\tt R\_RES}\cite{boonek}, an experiment--based
parametrization of low--mass quark pairs coupling to a virtual photon.
{\tt R\_RES} provides a reweighting factor 
which reproduces the experimentally observed resonant structures (R-factor) 
in $\gamma^* \ar q\bar q$ near  threshold .  When {\tt iresonance=1} is set,
the factor is applied to each  $q\bar q$ pair with $m(q\bar q) < 12$ GeV. 
One should be careful about the use of {\tt iresonance=1} in phase space 
regions where important low  $m(q\bar q)$ contributions are multiperipheral.
\subsubsection{Anomalous couplings} \label{anomalous}
\vsk
{\large\tt ianc}: this yes/no flag determines whether anomalous couplings
are used. If {\tt ianc=1} all of the following parameters must be set,
otherwise a fatal error will occur. They are defined in Ref.~\cite{ancoup}
\vsk
{\large\tt delz}, {\large\tt xf}, {\large\tt xz}, {\large\tt yf}, 
{\large\tt yz}, {\large\tt zz}
\vsk
\subsubsection{Cut selection} \label{cuts}
\vsk
{\large\tt icut}: this yes/no flag determines whether a predefined set of cuts
is implemented. If {\tt icut=1} all of the following parameters must be set,
otherwise a fatal error will occur.
\vsk
{\large\tt e\_min(e\_max)}: minimum (maximum) energy for the final state fermions.
Four values are required in the order in which the particles appear in the
Tables.
\vsk
{\large\tt rm\_min(rm\_max)}: minimum (maximum) invariant mass for pairs of final state fermions.
Six values are required.
They must be given in the following order:
    m(34), m(35), m(36), m(45), m(46), m(56). 
\vsk
{\large\tt pt\_min(pt\_max)}: minimum (maximum) transverse momentum for the final state fermions.
Four values are required.
\vsk
{\large\tt icos}: if {\tt icos=0} the limits on angles are given in degrees;
if {\tt icos=1} they refer to the cosine of the relevant angle.
Notice that the lower limit always refers to the smaller angle in degrees.
For example, if $0 \le \theta \le \pi$ and {\tt icos=1} then 
the following flags {\tt  thbeam\_min} and {\tt  thbeam\_max} must be set to 
1 and -1 respectively.
\vsk
{\large\tt thbeam\_min(thbeam\_max)}: minimum (maximum) angle of the final state fermions with respect to the
$e^+$ beam.
Four values are required. The accepted region is controlled by the following
flag {\tt iext}.
\vsk
{\large\tt iext}: if {\tt iext=0} the angle $\theta$ passes the cut
if  {\tt  theta\_min}$\le \theta \le${\tt  theta\_max} while 
if {\tt iext=1} the angle 
$\theta$ passes the cut if $\theta \le${\tt  theta\_min}
or {\tt  theta\_max}$\le \theta$ that is with {\tt iext=0} the interval
between the limits
is accepted while with {\tt iext=1} the acceptable region
is the one outside the limits. Four values are required, one for each final
fermion. This flag is useful for instance for processes with final 
particles lost in the beam pipe.
\vsk
{\large\tt thsep\_min(thsep\_max)}: minimum (maximum) angle between pairs of final state fermions.
Six values are required.
They must be given in the following order:
    (34), (35), (36), (45), (46), (56).
\vsk
{\large\tt ielost}: additional angular cuts for the process $e^+e^-\rightarrow
e^-e^+e^-e^+$. An electron is considered visible if it makes an angle larger 
than {\tt thetabeam} (see below)
with both beams. {\tt ielost=0} should be set when all four final particles are
required to be visible. {\tt ielost=1} requires instead only one invisible 
$e^-$ (the other three particles being outside the blind cone), 
{\tt ielost=2} only one invisible $e^+$. {\tt ielost=3} is for only a visible 
pair $e^-e^+$, {\tt ielost=4} requires the like-sign $e^-e^-$ pair to be 
visible, and {\tt ielost=5} the like-sign $e^+e^+$ pair visible. 
{\tt ielost=6} is for only a visible $e^+$, the other three particles being 
lost in the beam pipe, and finally {\tt ielost=7} is for only a visible $e^-$.
 If {\tt ielost>0}, the following input flag must
be set.
\vsk
{\large\tt thetabeam}: separation angle between visible and invisible 
$e^{\pm}$, to be specified if {\tt ielost>0}.  The value of {\tt thetabeam} must
be always expressed in degrees, independently on the flag {\tt icos}. 
\vsk
Additional cuts may be implemented by the user in \wphtwo (routine {\tt fxn.f})
 just after the line 
\begin{verbatim}
 * here define additional cuts 
\end{verbatim}
where a commented example is reported. The cuts must be specified using
the momenta of the final particles in the collider frame
{\tt p3(0:3)}, {\tt p4(0:3)}, {\tt p5(0:3)}, {\tt p6(0:3)} and the incoming
positron {\tt p1(0:3)} and incoming electron {\tt p2(0:3)}. 
They are assigned in the order in which the final state particles
appear in  Tables 1, 2 and 3.

\subsubsection{Distributions} \label{distributions}
\vsk
If distributions have to be automatically computed, they must first be
defined by the user in routine {\tt fxn.f}, after the line: 
\begin{verbatim}
* here define weighted distributions 
\end{verbatim}
where some commented examples of distributions are given.
The resulting cross sections corresponding to every
single bin will be stored in the file {\tt ABDIS.DAT}. Each line will contain
3 numbers: the value of the central point of the bin, the distribution for
the bin (i.e. the cross section divided by the width of the bin) and the 
statistical error. The next entries are unchanged from \wphonenosp.
\vsk
{\large\tt idistr}: this yes/no flag determines whether distributions are
desired.
If {\tt idistr=1} all of the following parameters must be set,
otherwise a fatal error will occur.
\vsk
{\large\tt ndistr}: the number of distributions to be generated
(limited by {\tt PARAMETER} {\tt ndismax} in routine {\tt wph.f}, 
whose present value is 50).
\vsk
For each required distribution the following set of parameters is mandatory
(The string $i$ must be replaced by the actual label of the distribution
{\tt 1}$\le i \le ${\tt ndistr}):
\vsk
{\large\tt nsubint($i$)}: number of sub-intervals with 
different binning (limited by {\tt PARAMETER} 
    {\tt nintmax} in routine {\tt wph.f}, whose present value is 10).
In normal plots one just uses one subinterval with a prescribed number
{\tt nbin\_number($1$)} (see below) of bins. If one chooses different
binning for different subintervals one can evidentiate in a plot particular
regions of interest.
\vsk
{\large\tt distr\_estrinf($i$)}: lower limits of each
  sub-interval (which coincide with the upper limit of the
          previous one) plus the upper limit of the last subint.
          {\tt nsubint($i$)+1} values are required.
\vsk
{\large\tt nbin\_number($i$)}: the number of bins for each sub-interval.
          {\tt nsubint($i$)} values are required
          (The total number of bins is limited by {\tt PARAMETER} 
    {\tt nbinmax} in routine {\tt wph.f}, whose present value is 500).
\vsk
\subsubsection{Unweighted event generation} \label{unweighted}
\vsk
This set of flags is relevant when generating unweighted
events for a single process or when preparing \infis\ for one--shot
generation.
\vsk
{\large\tt iflat}: this yes/no flag determines whether 
unweighted event generation is desired.
If {\tt iflat=1} all of the following parameters must be set,
otherwise a fatal error will occur.
{\tt iflat} must be equal to {\tt 1} when preparing \infis.
\vsk
{\large\tt scalemax}: factor by which the largest generated value of
the differential cross section is multiplied for the hit-or-miss selection.
This coefficient can be used to tune the efficiency of the hit-or-miss 
selection.
\vsk
{\large\tt istorvegas}: this yes/no flag determines whether 
{\tt VEGAS} data are stored after the last but one iteration in 
{\tt ABVEGAS.DAT} (or in {\tt ABVEGAS\_CC.DAT} and {\tt ABVEGAS\_NC.DAT} for 
NC+Higgs  and for Mixed processes when {\tt iccnc=3}). 
Stored {\tt VEGAS} data are necessary if one wants to rerun the program
  to generate again unweighted events. When the program is
  rerun using stored {\tt VEGAS} data, the maximum of the last
  iteration will be automatically used as the new maximum.
{\tt istorvegas} must be equal to {\tt 1} when preparing \infis.
\vsk
{\large\tt irepeat}: {\tt irepeat} has to be set to {\tt 0} for the first 
run. It has to be set to {\tt 1} if one wants to rerun the program exactly 
with the same input and grid starting from the last iteration. In this case the
same weighted points will be reproduced unless {\tt scalemax} is varied.
In the first two cases the number of attempts is fixed but not that of 
generated unweighted events.
  {\tt irepeat=2}  has to be chosen if one wants  to rerun with the same input
   and grid as  before, but letting the program run until a requested number of
  events {\tt nflevts} is reached. For both cases {\tt irepeat=1} and {\tt 2}
  one might of course vary   {\tt scalemax}, {\tt ijetset} and {\tt istormom}
  (see below as to the last two entries)
  with respect to the first run with {\tt irepeat=0}.
{\tt irepeat} must be equal to {\tt 0} when preparing \infis.
\vsk
{\large\tt nflevts}: the number of unweighted events to be generated
if {\tt irepeat=2}.
\vsk
{\large\tt istormom}: this yes/no flag determines whether the momenta of the
generated unweighted events have to be stored or not.
If {\tt istormom=1}, the momenta of the unweighted events are generally 
written in single precision  in the {\tt ABMOM.DAT}. 
In the particular case of Mixed (when {\tt iccnc=3}) or 
 NC+Higgs processes, the {\tt ABMOM\_SIGN.DAT} and 
{\tt ABMOM\_BACK.DAT} files will be produced.
{\tt ABMOM\_SIGN.DAT} is used for CC or Higgs events, {\tt ABMOM\_BACK.DAT} 
for NC events.
\vsk
As to the entries relative to this subsection, no substantial change has been
introduced  with respect to the previous version of \wphnosp, with the following
exception.
In the new version the number of iterations required for the unweighted event 
generation is not constrained anymore to be {\tt itmx=2}, but can be
freely chosen as in all other cases.\parno
Every unweighted event is passed to the standard {\tt COMMON/HEPEVT}.
\vsk
{\large\tt ijetset}: this yes/no flag determines whether the program should
be interfaced to \pytnosp . If {\tt ijetset=1} the subroutine {\tt AB\_LU4FRM} 
is called for interfacing \pyt (or {\tt Hadgen} for $m (q\bar q) < 2$ GeV), 
otherwise {\tt HEPEVT} is still filled but the interface to \pyt is not called.
\vsk
\subsubsection{Integration} \label{integration}
\vsk
Unchanged from \wphonenosp.
\vsk
{\large\tt acc}: the integration accuracy. When accuracy {\tt acc} is reached 
the program stops.
\vsk
{\large\tt iterm}: this yes/no flag determines whether {\tt VEGAS} has to 
adapt its integration grid by performing thermalization iterations 
whose outcome will not be used in the final result.
\vsk
{\large\tt ncall\_term}: the maximum number of points for each grid refinement
during thermalization.
\vsk
{\large\tt itmx\_term}: the number of times the grid is adapted before 
starting the actual integration
(limited by {\tt PARAMETER} {\tt nitmax} in routine {\tt wph.f}, whose present value is 10).
\vsk
{\large\tt ncall}: the maximum number of points for each iteration of the 
actual integration. 
\vsk
{\large\tt itmx}: the maximum number of iterations used to evaluate the 
integral
(limited by {\tt PARAMETER} {\tt nitmax} in routine {\tt wph.f}, whose present value is 10).
A value among 3 and 5 is normally the best choice.
   If higher precision is requested it is usually more convenient to 
  increase {\tt ncall} rather than   {\tt itmx}.

\vsk
{\tt VEGAS} will in general use a number of {\tt ncall\_term} and {\tt ncall} 
lower than the input ones. The actual value is written in output, where
also the number of points which survive all the cuts ({\tt effective ncall}) 
is reported. 
    
As a final remark about the choice of these parameters, one must be aware
    of the fact that final results with a $\chi^2$ much greater than the 
   number of  iterations  are not to be trusted. When this happens, one has to
   increase {\tt ncall}. 

\vsk
\subsubsection{Higgs} \label{higgs}
\vsk
All the final states in the NC+Higgs group are also included in the NC class.
The difference is the following.
The matrix elements present in the NC group do not include the Higgs boson
as an intermediate state and all particles can be either massive
({\tt imass=1}) or massless ({\tt imass=0}). 
The class of processes given by {\tt iproc $\geq$ 33} includes instead the Higgs boson
as an intermediate particle and are optimized
for Higgs searches. In these processes, all initial and final particles 
are massless, except those which might come from the Higgs decay, i.e. 
massive $b$'s, $c$'s or $\tau$'s. In Table 3 the complete list of these
channels is shown. For each channel with no identical particles in the
final state, the order in which the final particles appear is such that the 
first $f\bar f$ pair is the massive one (i.e. the pair that could come from 
the Higgs decay). For processes with identical particles, like 
$e^+e^-\rightarrow b\bar bb\bar b$, the final state is completely 
massive.
\vsk
{\large\tt rmb (rmc, rmtau)}: sets the $b$ ($c$, $\tau$) mass in both phase 
space and matrix element for processes where the Higgs could decay into 
$b\bar b$ ($c\bar c$, $\tau^-\tau^+$).
\vsk
{\large\tt rmbrun (rmcrun)}: sets the $b$ ($c$) mass in the Higgs 
coupling, which might be different from {\tt rmb (\tt rmc)}.
\vsk
The processes listed in Table 3 can receive contributions either from 
the SM Higgs or from MSSM neutral Higgses (the CP-odd $A$ and the lightest 
CP-even $h$). 
\vsk
{\large\tt isusy}: {\tt isusy=0} corresponds to SM Higgs whereas {\tt isusy=1}
to the MSSM Higgs sector.
\vsk
Also for these channels the user can select subsets of Feynman diagrams.
\vsk
{\large\tt icch}: if {\tt icch=1} the subset is the Higgs signal. If 
{\tt icch=2} the subset is the Higgs background (i.e. the complete set
of Feynman diagrams without the Higgs). If {\tt icch=3} the full process
is computed (Higgs+Background+interference).
\vsk
Moreover, when computing the pure Higgs signal ({\tt icch=1}) for the three
processes $b\bar bb\bar b$, $c\bar cc\bar c$ and $\tau^-\tau^+\tau^-\tau^+$, 
in the new version of \wph the user can also choose:
\vsk
{\large\tt iha}: for SM Higgs, {\tt iha=1} gives the complete amplitude, 
while {\tt iha=2} computes only the Feynman diagrams 
corresponding to $h$Z production and decay. For MSSM Higgs, {\tt iha=1} gives
as before the complete amplitude. {\tt iha=2} selects only the
diagrams which describe single $h$ production ({\it i.e.} diagrams which 
are singly-resonant on the $h$ mass), {\tt iha=3} the diagrams for
single $A$ production ({\it i.e.} diagrams which 
are singly-resonant on the $A$ mass).
{\tt iha=4} computes the subset for $h$Z production,
{\tt iha=5} gives the $hA$ associate pair production and finally {\tt iha=6}
generates the sum of the two contributions $h$Z+$hA$.  
\vsk
{\large\tt rmh}: the mass of the SM Higgs or the mass of the 
lightest MSSM CP-even Higgs.
\vsk
{\large\tt irmhcomp}: this flag allows the user either
to choose arbitrary values for the 
lightest MSSM CP-even Higgs {\tt rmh} ({\tt irmhcomp=0}) or to compute
{\tt rmh} using standard formulae ({\tt irmhcomp=1}). In the first case, one
should specify the chosen {\tt rmh} value, followed by
\vsk
{\large\tt rma}: the MSSM CP-odd Higgs mass
\vsk
{\large\tt tgb}: the value of $tg(\beta)$, the ratio between the two vacuum 
expectation values.
\vsk
In the latter case ({\tt irmhcomp=1}), given {\tt rma} and {\tt tgb} as input,
the value of {\tt rmh} is automatically computed. The calculation is
controlled by:
\vsk
{\large\tt iloop}: if {\tt iloop=1} the lightest CP-even Higgs mass is
determined at 1-loop  (see \wphone). If {\tt iloop=2}, {\tt rmh}  
is derived according to the 2-loop RG improved effective potential 
results given in Ref.~\cite{2loophiggs}.
\vsk
In the {\tt iloop=2} case, the user must specify one of the 
following possible scenarios:
\vsk
{\large\tt imixing}: {\tt imixing=1} corresponds to the {\tt no mixing} case
($A_t$=$A_b$=0 and $|\mu |\ll M_S$ where $M_S$ is the SUSY scale),
 {\tt imixing=2} corresponds to {\tt maximal mixing} 
($A_t$=$A_b$=$\sqrt{6}M_S$ and $|\mu |\ll M_S$), and {\tt imixing=3}
corresponds to {\tt typical mixing} ($A_t$=$A_b$=-$\mu$).
Finally, {\tt imixing=4} allows the user to select arbitrary values for 
\vsk
{\large\tt At}: the trilinear soft susy breaking term $A_t$;
\vsk
{\large\tt Ab}: the trilinear soft susy breaking term $A_b$.
\vsk
{\large\tt rmyou}: the SUSY Higgs mass parameter $\mu$.
\vsk

\vskip 1cm
\noindent
   In addition to the parameters in input, other physical quantities
   are fixed in the routine {\tt wph.f}  by a {\tt DATA} 
   statement, where
\vsk 
{\tt rmw, rmz, rmt, rmb, rmc, rms, rmu, rmd, rme, rmmu, rmtau} are 
respectively the  
W, Z, $t$, $b$, $c$, $s$, $u$, $d$, $e$, $\mu$ and $\tau$ masses;
\parno
{\tt rmb\_run } is the quark $b$ mass used for the Higgs coupling;
\parno
{\tt gamw}, {\tt gamz}, {\tt gamh} are the total W, Z and Higgs width;
\parno
{\tt gf } is the Fermi coupling constant;
{\tt alfainv } is $1/\alpha_{QED}$ at the appropriate scale; {\tt alfas\_cc }
and {\tt alfas\_nc } are $\alpha_s(M_W)$ and $\alpha_s(M_Z)$;
\parno
{\tt s2w } is the Weinberg $\sin^2(\theta_W)$ and {\tt rmsus} the SUSY scale;
\parno
{\tt vud, vus, vcd, vcs, vcb} are the CKM matrix elements.
\vsk
\subsection{Multiple process case: {\tt ionesh=1}} \label{ionesh1}
\vsk
{\tt ionesh=1} can be used only for massive processes ({\tt imass=1})
\vsk
If {\tt ionesh=1} the following flags are required:
\vsk
{\large \tt e\_cm}: the centre of mass energy. This does not need to be 
equal to the energy at which the \infis\ have been produced.
 \vsk
{\large \tt scalemax}:  factor by which the largest generated value of
the differential cross section is multiplied for the hit-or-miss selection.
This coefficient can be used to tune the efficiency of the hit-or-miss 
selection. It prevents generating too many events with weight larger
than the maximum stored in the \infi. In our experience {\tt scalemax=1.1}
is usually a good choice.
\vsk
{\large \tt nunwevts}: number of unweighted events to be generated.
\vsk
{\large \tt iallme}: this yes/no flag determines whether, for every unweighted
event, the matrix element squared for all applicable subsets of Feynman 
diagrams are computed. If  {\tt iallme=1} the 
partial results are stored in the variable {\tt SQMEL(19)} passed to the 
{\tt COMMON/WPHSME}.
\vsk
{\large \tt istormom}: this yes/no flag determines whether the momenta of the
generated unweighted events have to be stored or not.
If {\tt istormom=1}, the momenta of the unweighted events are 
written in {\tt ABMOM.DAT}.\parno
\vsk
{\large \tt ijetset}: this yes/no flag determines whether the program should
be interfaced to \pyt. If {\tt ijetset=1} the subroutine {\tt AB\_LU4FRM} 
is called for interfacing \pyt (or {\tt Hadgen} for $m (q\bar q) < 2$ GeV), 
otherwise {\tt HEPEVT} is still filled but 
the interface to \pyt is not called.\parno
\vsk
{\large \tt iallch}: if {\tt iallch=0} only those processes for which an \infi\ 
is present in the input list are generated.  If {\tt iallch=1}, 
for each process for which an \infi\ is present in the input list, the full set of processes
which have the same {\tt iproc} will be included in the generation. In this
case, the processes for which \infis\ are present  must have {\tt ich=1}
and {\tt ickm=0}. The subsequent flag {\tt ickm}, which has to be specified
 in the one-shot input,  determines then whether the full set must take
into account CKM matrix effects. 
\vsk
{\large \tt ickm}: this flag determines whether the 
Cabibbo--Kobayashi--Maskawa matrix is taken into account ({\tt ickm=1}) or
it is taken equal to the unit matrix ({\tt ickm=0}). 
Notice that for {\tt ickm=1} the number of channels in each CC group increases.
This flag is effective only if {\tt iallch=1}. If {\tt iallch=0}, only
the processes of the input files are generated, and their {\tt ickm} flags
determine eventual CKM effects.
\vsk
{\large \tt nfiles}: the number of \infis\ to be used for one--shot generation
(limited by {\tt PARAMETER} {\tt nmaxproc}, whose present value is 150).

\vsk
The {\tt nfiles} line must be immediately followed by as many filenames
(with complete relative path), each on a separate line, as indicated by the {\tt nfiles} value.
All other input variables are read directly from the \infis.
\vsk

\section{Conclusions}\label{conclusion}

We have described version 2.0 of \wphnosp , a fully massive MC program and
unweighted event generator
which computes all Standard Model processes with four fermions in the
final 
state at $e^+ e^-$ colliders. Thanks to the new features like fermion
masses,
the IFL gauge restoring scheme and new phase space mappings
\wph has been extended to all regions of phace space,
including 
kinematical configurations dominated by small momentum transfer and
small invariant masses
like single W, single Z, Z$\gamma ^*$ and $\gamma ^*\gamma ^*$
processes.
Special attention has been devoted to QED effects, which have a large
numerical
impact, with new options for the description of Initial State Radiation
and
of the scale dependence of the electromagnetic coupling.
The program is now better suited to generate large samples of fully
simulated
events since it can produce unweighted events for any user selected
subset of $4f$ final states in a single run.
\vskip .5cm \noindent
The program is available from {\tt
http://www.to.infn.it/$^\sim$ballestr/wphact/2.0/}
together with fully commented examples of input files.
All future improvements and bug fixes will be distributed through the
site.

\section*{Acknowledgments}
We gratefully acknowledge useful discussions and comparisons with Giampiero
Passarino and Roberto Pittau. A. B. wishes to thank 
P. Bambade, M. Boonekamp, R. Chierici, F. Cossutti, E. Migliore  and many 
other members of the Delphi Collaboration for advice, suggestions and  
discussions. 
We are grateful to F. Cossutti also for providing the code used to interface
{\tt PHOTOS}.

\vfill\eject

\section*{Test run }

We report here  an example of  test run  for one of the single Z cross sections
presented in \fig{singleZq}.
\vsk
\parno
{\bf Input}
\parno
\wph requires from {\tt standard input} the name of the file from
which all flags are read.  The content of this file for  this case is 
reported below. In the distribution of \wph a complete and fully commented
example of input file is included. The comments, which remind the meaning
and the available choices of the various flags, are not reproduced here,
but they can be left in place in actual computations as they are ignored
by the program. 

\begin{quote}{\footnotesize \begin{verbatim}
 ionesh            0
 e_cm              199.5d0     
 rmw               80.40d0     
 iproc             26
 ickm              0
 ich               1
 imass             1
 icc3              0
 izz               0
 izg               0
 izg34             0
 izg56             0
 igamgams          0
 inc08             0
 imulper           0
 igamgam           0
 itch              0
 itchnomp          0
 ips_cc            3
 ips_nc            3
 ismallangle       1
 ismallangle_ee    3
 ialftsw           0
 ialfar            1
 iresonance        0
 isr               2
 ibeam             0
 icoul             0
 istrcor           1
 ianc              0
 ipr               1
 iswgcomp          1
 igwcomp           1
 igzcomp           1
 ighcomp           1
 icut              1
 e_min        0.d0   3.d0   0.d0    0.d0
 e_max        5000.d0   5000.d0   5000.d0   5000.d0
 rm_min       0.d0   0.d0   0.d0   0.d0   0.d0   60.d0     
 rm_max       5000.d0   5000.d0   5000.d0   5000.d0   5000.d0   5000.d0     
 pt_min       0.d0   0.d0   0.d0   0.d0
 pt_max       5000.d0   5000.d0   5000.d0   5000.d0     
 icos              0
 iext              0   0   0   0
 thbeam_min   168.d0   60.d0   0.d0   0.d0
 thbeam_max   180.d0   168.d0   180.d0   180.d0     
 thsep_min    0.d0   0.d0   0.d0   0.d0   0.d0   0.d0
 thsep_max    180.d0   180.d0   180.d0   180.d0   180.d0   180.d0     
 idistr            0
 iflat             0
 acc               0.d0
 iterm             1
 ncall_term        200000
 itmx_term         4
 ncall             4000000
 itmx              4
\end{verbatim}}\end{quote}

\parno
{\bf Output}
\parno
In the output file produced by \wph one will recover all relevant 
input flags  (not reported here) followed (for {\tt ionesh=0})
by the process name, a list of physical parameters, the result
of the thermalization iterations and finally the result of the  
iterations for the actual integration.

\begin{quote}{\footnotesize \begin{verbatim}
  
 NC48 )  e-(p3) e+(p4) mu-(p5) mu+(p6)
  
 INPUT
 cm energy             =          0.1995000D+03 GeV
  
 DATA
 Z mass                =          0.9118700D+02 GeV
 W mass                =          0.8040000D+02 GeV
 electron mass         =          0.5109991D-03 GeV
 mu mass               =          0.1056583D+00 GeV
 tau mass              =          0.1777000D+01 GeV
 up mass               =          0.5000000D-02 GeV
 down mass             =          0.1000000D-01 GeV
 charm mass            =          0.1300000D+01 GeV
 strange mass          =          0.2000000D+00 GeV
 top mass              =          0.1750000D+03 GeV
 bottom mass           =          0.4800000D+01 GeV
 Gf                    =          0.1166390D-04 GeV-2
 alfas_nc              =          0.1230000D+00
  
 DERIVED QUANTITIES
 W width               =          0.2101131D+01 GeV
 Z width               =          0.2506693D+01 GeV
 s2w                   =          0.2225969D+00
 1/alfa_em             =          0.1323609D+03
 1/alfa_em(q**2=0)     =          0.1370360D+03
  
 OPTIONS
 Z s-dependent width
 QEDPS
 Naive QCD corrections included
 Single resonant (56)"NC" phase space
------------------------------------------------------
  
  
 Thermalization

 input parameters for vegas:  ndim=  7  ncall=     156250.
                              it=    1  itmx=    4

 iteration no.  1:            effective ncall=      60114
 iteration no.  1: integral = 0.2211926E-01+/-  0.37E-03
 all iterations:   integral = 0.2211926E-01+/- 0.367E-03 chi**2/it'n = 0.00E+00

 iteration no.  2:            effective ncall=      93901
 iteration no.  2: integral = 0.2232726E-01+/-  0.15E-03
 all iterations:   integral = 0.2229726E-01+/- 0.139E-03 chi**2/it'n = 0.27    

 iteration no.  3:            effective ncall=      95374
 iteration no.  3: integral = 0.2234192E-01+/-  0.14E-03
 all iterations:   integral = 0.2231946E-01+/- 0.987E-04 chi**2/it'n = 0.16    

 iteration no.  4:            effective ncall=      93411
 iteration no.  4: integral = 0.2228780E-01+/-  0.11E-03
 all iterations:   integral = 0.2230523E-01+/- 0.733E-04 chi**2/it'n = 0.12    
  
 NC process

 input parameters for vegas:  ndim=  7  ncall=    3294172.
                              it=    1  itmx=    4

 iteration no.  1:            effective ncall=    2027248
 iteration no.  1: integral = 0.2237196E-01+/-  0.21E-04
 all iterations:   integral = 0.2237196E-01+/- 0.208E-04 chi**2/it'n = 0.00E+00

 iteration no.  2:            effective ncall=    2012237
 iteration no.  2: integral = 0.2238736E-01+/-  0.20E-04
 all iterations:   integral = 0.2237985E-01+/- 0.145E-04 chi**2/it'n = 0.28    

 iteration no.  3:            effective ncall=    1985637
 iteration no.  3: integral = 0.2238947E-01+/-  0.24E-04
 all iterations:   integral = 0.2238239E-01+/- 0.124E-04 chi**2/it'n = 0.20    

 iteration no.  4:            effective ncall=    1964560
 iteration no.  4: integral = 0.2238508E-01+/-  0.21E-04
 all iterations:   integral = 0.2238308E-01+/- 0.107E-04 chi**2/it'n = 0.14    
  
------------------------------------------------------
  
 Sigma = 0.2238308D-01 +/- 0.107D-04 (pb)
\end{verbatim}}\end{quote}
\vfill\eject

\begin{table}[hbt]\centering
\begin{tabular}{|c|c|c l c l|}
\hline 
\multicolumn{6}{|c|}{} \\
\multicolumn{6}{|c|}{\bf CC}\\
\hline
process type&iproc& ich & final state& ich & final state \\
\hline
\hline
CC9&1&1&$\mu^-$ $\bar\nu_\mu$ $\nu_\tau$ $\tau^+$  
 &2&$\mu^+$  $\nu_\mu$  $\bar\nu_\tau$  $\tau^-$\\
\hline
CC18&2&1&$e^-$  $\bar\nu_e$  $\nu_\mu$  $\mu^+$ 
 &3&$e^+$  $\nu_e$  $\bar\nu_\mu$  $\mu^-$ \\
& &2&$e^-$  $\bar\nu_e$  $\nu_\tau$  $\tau^+$ 
 &4&$e^+$  $\nu_e$  $\bar\nu_\tau$  $\tau^-$ \\
\hline
CC10&3&1&$\mu^-$  $\bar\nu_\mu$  $u$  $\bar d$ 
 &11&$\mu^-$  $\bar\nu_\mu$  $c$  $\bar b$ \\
& &2&$\mu^-$  $\bar\nu_\mu$  $c$  $\bar s$ 
 &12&$\mu^+$  $\nu_\mu$  $\bar u$  $s$ \\
& &3&$\mu^+$  $\nu_\mu$  $\bar u$  $d$ 
 &13&$\mu^+$  $\nu_\mu$  $\bar c$  $d$ \\
& &4&$\mu^+$  $\nu_\mu$  $\bar c$  $s$ 
 &14&$\mu^+$  $\nu_\mu$  $\bar c$  $b$ \\
& &5&$\tau^-$  $\bar\nu_\tau$  $u$  $\bar d$ 
 &15&$\tau^-$  $\bar\nu_\tau$  $u$  $\bar s$ \\
& &6&$\tau^-$  $\bar\nu_\tau$  $c$  $\bar s$ 
 &16&$\tau^-$  $\bar\nu_\tau$  $c$  $\bar d$ \\
 & &7&$\tau^+$  $\nu_\tau$  $\bar u$  $d$ 
 &17&$\tau^-$  $\bar\nu_\tau$  $c$  $\bar b$ \\
& &8&$\tau^+$  $\nu_\tau$  $\bar c$  $s$ 
 &18&$\tau^+$  $\nu_\tau$  $\bar u$  $s$ \\
 & &9&$\mu^-$  $\bar\nu_\mu$  $u$  $\bar s$ 
 &19&$\tau^+$  $\nu_\tau$  $\bar c$  $d$ \\
& &10&$\mu^-$  $\bar\nu_\mu$  $c$  $\bar d$ 
 &20&$\tau^+$  $\nu_\tau$  $\bar c$  $b$ \\
\hline
CC20&4&1&$e^-$  $\bar\nu_e$  $u$  $\bar d$ 
 &6&$e^-$  $\bar\nu_e$  $c$  $\bar d$ \\
& &2&$e^-$  $\bar\nu_e$  $c$  $\bar s$ 
 &7&$e^-$  $\bar\nu_e$  $c$  $\bar b$ \\
& &3&$e^+$  $\nu_e$  $\bar u$  $d$ 
 &8&$e^+$  $\nu_e$  $\bar u$  $s$ \\
& &4&$e^+$  $\nu_e$  $\bar c$  $s$ 
 &9&$e^+$  $\nu_e$  $\bar c$  $d$ \\
& &5&$e^-$  $\bar\nu_e$  $u$  $\bar s$ 
 &10&$e^+$  $\nu_e$  $\bar c$  $b$ \\
\hline
CC11&5&1&$s$  $\bar c$  $u$  $\bar d$ 
 &13&$d$  $\bar c$  $c$  $\bar s$ \\
& &2&$\bar s$  $c$  $\bar u$  $d$ 
 &14&$b$  $\bar c$  $c$  $\bar s$\\
& &3&$s$  $\bar c$  $u$  $\bar s$ 
 &15&$d$  $\bar c$  $u$  $\bar s$\\
& &4&$d$  $\bar c$  $u$  $\bar d$ 
 &16&$b$  $\bar c$  $u$  $\bar s$\\
& &5&$b$  $\bar c$  $u$  $\bar d$ 
 &17&$\bar d$  $c$  $\bar u$ $s$\\
& &6&$\bar s$  $c$  $\bar u$  $s$ 
 &18&$\bar b$  $c$  $\bar u$  $s$\\
& &7&$\bar d$  $c$  $\bar u$  $d$ 
 &19&$d$  $\bar c$  $c$  $\bar b$\\
& &8&$\bar b$  $c$  $\bar u$  $d$ 
 &20&$b$  $\bar c$  $c$  $\bar d$\\
& &9&$d$  $\bar u$  $u$  $\bar s$ 
 &21&$s$  $\bar u$  $u$  $\bar s \ ^{\ **}$ \\
& &10&$s$  $\bar u$  $u$  $\bar d$ 
 &22&$d$  $\bar c$  $c$  $\bar d \ ^{\ **}$ \\
& &11&$s$  $\bar c$  $c$  $\bar d$ 
 &23&$b$  $\bar c$  $c$  $\bar b \ ^{\ **}$  \\
& &12&$s$  $\bar c$  $c$  $\bar b$ 
 & &       \\
\hline
\hline
\multicolumn{6}{|c|}{} \\
\multicolumn{6}{|c|}{\bf MIX}\\
\hline
process type& iproc&ich & final state& ich & final state \\
\hline
\hline
MIX19&6&1&$\mu^-$  $\mu^+$  $\nu_\mu$  $\bar\nu_\mu$ 
 &2&$\tau^-$  $\tau^+$  $\nu_\tau$  $\bar\nu_\tau$ \\
\hline
MIX56&7&1&$e^-$  $e^+$  $\nu_e$  $\bar\nu_e$ 
 &&\\
\hline
MIX43&8&1&$d$  $\bar d$  $u$  $\bar u$ 
 &2&$s$  $\bar s$  $c$  $\bar c$ \\
\hline
\end {tabular}
\caption {CC and MIX processes. They can be computed with massive particles
({\tt imass=1}) or in the massless approximation ({\tt imass=0}).
The processes indicated by $^{**}$  correspond only to the CC contributions 
induced by CKM mixing. 
Their  normal NC contributions have to be evaluated 
with {\tt iproc=24}, {\tt ich=1,2,4} respectively. }
\end {table}

\begin{table}[hbt]\centering
\begin{tabular}{|c|c|c l c l|}
\hline 
\multicolumn{6}{|c|}{} \\
\multicolumn{6}{|c|}{\bf NC}\\
\hline
process type& iproc&ich & final state& ich & final state \\
\hline
\hline
NC6&9&1&$\nu_\mu$  $\bar\nu_\mu$  $\nu_\tau$  $\bar\nu_\tau$ 
&&\\
\hline
NC12&10&1&$\nu_\mu$  $\bar\nu_\mu$  $\nu_e$  $\bar\nu_e$
  &2&$\nu_\tau$  $\bar\nu_\tau$  $\nu_e$  $\bar\nu_e$\\
\hline
NC12&11&1&$\nu_\mu$  $\bar\nu_\mu$  $\nu_\mu$  $\bar\nu_\mu$
  &2&$\nu_\tau$  $\bar\nu_\tau$  $\nu_\tau$  $\bar\nu_\tau$\\
\hline
NC36&12&1&$\nu_e$  $\bar\nu_e$  $\nu_e$  $\bar\nu_e$
&&\\
\hline
NC10&13&1&$u$  $\bar u$  $\nu_\mu$  $\bar\nu_\mu$
  &3&$c$  $\bar c$  $\nu_\mu$  $\bar\nu_\mu$\\
&  &2&$u$  $\bar u$  $\nu_\tau$  $\bar\nu_\tau$
  &4&$c$  $\bar c$  $\nu_\tau$  $\bar\nu_\tau$\\
\hline
NC19&14&1&$u$  $\bar u$  $\nu_e$  $\bar\nu_e$
  &2&$c$  $\bar c$  $\nu_e$  $\bar\nu_e$\\
\hline
NC64&15&1&$u$  $\bar u$  $u$  $\bar u$
  &2&$c$  $\bar c$  $c$  $\bar c$\\
\hline
NC32&16&1&$u$  $\bar u$  $c$  $\bar c$
  &&\\
\hline
NC10&17&1&$\mu^-$  $\mu^+$  $\nu_\tau$  $\bar\nu_\tau$
  &2&$\tau^-$  $\tau^+$  $\nu_\mu$  $\bar\nu_\mu$\\
\hline
NC20&18&1&$e^-$  $e^+$  $\nu_\mu$  $\bar\nu_\mu$
  &2&$e^-$  $e^+$  $\nu_\tau$  $\bar\nu_\tau$\\
\hline
NC19&19&1& $\mu^-$  $\mu^+$  $\nu_e$  $\bar\nu_e$ 
  &2& $\tau^-$  $\tau^+$  $\nu_e$  $\bar\nu_e$ \\
\hline
NC24&20&1& $\mu^-$  $\mu^+$  $u$  $\bar u$ 
  &3& $\tau^-$  $\tau^+$  $u$  $\bar u$ \\
&  &2& $\mu^-$  $\mu^+$  $c$  $\bar c$ 
  &4& $\tau^-$  $\tau^+$  $c$  $\bar c$ \\
\hline
NC48&21&1& $e^-$  $e^+$  $u$  $\bar u$ 
  &2& $e^-$  $e^+$  $c$  $\bar c$ \\
\hline
NC19&22&1& $d$  $\bar d$  $\nu_e$  $\bar\nu_e$ 
  &3& $b$  $\bar b$  $\nu_e$  $\bar\nu_e$ \\
&  &2& $s$  $\bar s$  $\nu_e$  $\bar\nu_e$ 
  &&\\
\hline
NC10&23&1& $d$  $\bar d$  $\nu_\mu$  $\bar\nu_\mu$ 
  &4& $d$  $\bar d$  $\nu_\tau$  $\bar\nu_\tau$ \\
&  &2& $s$  $\bar s$  $\nu_\mu$  $\bar\nu_\mu$ 
  &5& $s$  $\bar s$  $\nu_\tau$  $\bar\nu_\tau$ \\
&  &3& $b$  $\bar b$  $\nu_\mu$  $\bar\nu_\mu$ 
  &6& $b$  $\bar b$  $\nu_\tau$  $\bar\nu_\tau$ \\
\hline
NC32&24&1& $s$  $\bar s$  $u$  $\bar u$ 
  &3& $b$  $\bar b$  $u$  $\bar u$ \\
&  &2& $d$  $\bar d$  $c$  $\bar c$ 
  &4& $b$  $\bar b$  $c$  $\bar c$ \\
\hline
NC24&25&1& $\mu^-$  $\mu^+$  $\tau^-$  $\tau^+$ 
  &&\\
\hline
NC48&26&1& $e^-$  $e^+$  $\mu^-$  $\mu^+$ 
  &2& $e^-$  $e^+$  $\tau^-$  $\tau^+$ \\
\hline
NC48&27&1& $\mu^-$  $\mu^+$  $\mu^-$  $\mu^+$ 
  &2& $\tau^-$  $\tau^+$  $\tau^-$  $\tau^+$ \\
\hline
NC144&28&1& $e^-$  $e^+$  $e^-$  $e^+$ 
  &&\\
\hline
NC64&29&1& $d$  $\bar d$  $d$  $\bar d$ 
  &3& $b$  $\bar b$  $b$  $\bar b$ \\
&  &2& $s$  $\bar s$  $s$  $\bar s$ 
  &&\\ 
\hline
NC48&30&1& $e^-$  $e^+$  $d$  $\bar d$ 
  &3& $e^-$  $e^+$  $b$  $\bar b$ \\
&  &2& $e^-$  $e^+$  $s$  $\bar s$ 
  &&\\ 
\hline
NC24&31&1& $\mu^-$  $\mu^+$  $d$  $\bar d$ 
  &4& $\tau^-$  $\tau^+$  $d$  $\bar d$ \\
&  &2& $\mu^-$  $\mu^+$  $s$  $\bar s$ 
  &5& $\tau^-$  $\tau^+$  $s$  $\bar s$ \\
&  &3& $\mu^-$  $\mu^+$  $b$  $\bar b$ 
  &6& $\tau^-$  $\tau^+$  $b$  $\bar b$ \\
\hline
NC32&32&1& $d$  $\bar d$  $s$  $\bar s$ 
  &3& $s$  $\bar s$  $b$  $\bar b$ \\
&  &2& $d$  $\bar d$  $b$  $\bar b$ 
  &&\\
\hline
\end {tabular}
\caption {NC processes. They can be computed with massive particles
({\tt imass=1}) or in the massless approximation ({\tt imass=0}). }
\end {table}

\vfill\eject

\begin{table}[hbt]\centering
\begin{tabular}{|c|c|c l c l|}
\hline
\multicolumn{6}{|c|}{} \\
\multicolumn{6}{|c|}{\bf NC+HIGGS}\\
\hline
process type& iproc& ich & final state& ich & final state \\
\hline
\hline
NC21&33&1& $b$  $\bar b$  $\nu_e$  $\bar\nu_e$ 
  &&\\
\hline
NC11&34&1& $b$  $\bar b$  $\nu_\mu$  $\bar\nu_\mu$ 
  &2& $b$  $\bar b$  $\nu_\tau$  $\bar\nu_\tau$ \\
\hline
NC33&35&1& $b$  $\bar b$  $u$  $\bar u$ 
  &2& $b$  $\bar b$  $c$  $\bar c$ \\
\hline
NC25&36&1& $b$  $\bar b$  $\mu^-$  $\mu^+$ 
  &2& $b$  $\bar b$  $\tau^-$  $\tau^+$ \\
\hline
NC50&37&1& $b$  $\bar b$  $e^-$  $e^+$ 
  && \\
\hline
NC33&38&1& $b$  $\bar b$  $d$  $\bar d$ 
  &2& $b$  $\bar b$  $s$  $\bar s$ \\
\hline
NC84&39&1& $b$ $\bar b$  $b$  $\bar b$ 
  &&\\
\hline
NC21&40&1& $c$  $\bar c$  $\nu_e$  $\bar\nu_e$ 
  &&\\
\hline
NC11&41&1& $c$  $\bar c$  $\nu_\mu$  $\bar\nu_\mu$ 
  &2& $c$  $\bar c$  $\nu_\tau$  $\bar\nu_\tau$ \\
\hline
NC33&42&1& $c$  $\bar c$  $u$  $\bar u$ 
  && \\
\hline
NC84&43&1& $c$  $\bar c$  $c$  $\bar c$ 
  && \\ 
\hline
NC25&44&1& $c$  $\bar c$  $\mu^-$  $\mu^+$ 
  &2& $c$  $\bar c$  $\tau^-$  $\tau^+$ \\
\hline
NC50&45&1& $c$  $\bar c$  $e^-$  $e^+$ 
  && \\
\hline
NC33&46&1& $c$  $\bar c$  $d$  $\bar d$ 
  &3& $c$  $\bar c$  $b$  $\bar b$ \\
NC44&  &2& $c$  $\bar c$  $s$  $\bar s$ 
  &&\\ 
\hline
NC21&47&1& $\tau^-$  $\tau^+$  $\nu_e$  $\bar\nu_e$ 
  &&\\
\hline
NC11/NC20&48&1& $\tau^-$  $\tau^+$  $\nu_\mu$  $\bar\nu_\mu$ 
  &2& $\tau^-$  $\tau^+$  $\nu_\tau$  $\bar\nu_\tau$ \\
\hline
NC25&49&1& $\tau^-$  $\tau^+$  $u$  $\bar u$
  &2& $\tau^-$  $\tau^+$  $c$  $\bar c$ \\ 
\hline
NC25&50&1& $\tau^-$  $\tau^+$  $\mu^-$  $\mu^+$
  && \\ 
\hline
NC68&51&1& $\tau^-$  $\tau^+$  $\tau^-$  $\tau^+$
  && \\ 
\hline
NC50&52&1& $\tau^-$  $\tau^+$  $e^-$  $e^+$ 
  && \\
\hline
NC25&53&1& $\tau^-$  $\tau^+$  $d$  $\bar d$ 
  &3& $\tau^-$  $\tau^+$  $b$  $\bar b$ \\
&  &2& $\tau^-$  $\tau^+$  $s$  $\bar s$ 
  &&\\
\hline
\end {tabular}
\caption {NC processes which take into account Higgs contributions.
 They require {\tt imass=0} and have by default 
 massive $b$'s in processes 33-39, 
 massive $c$'s in processes 40-46 and 
 massive $\tau$'s in processes 47-53. All other 
external particles are massless.}
\end {table}


\begin{thebibliography}{1}

\bibitem{yr} {\it Physics at LEP2}, G.~Altarelli, T.~Sj\"ostrand and 
F.~Zwirner eds., CERN 96-01 (1996).

\bibitem{yr2k} 
{\it Reports of the Working Groups on Precision Calculations
for LEP2 Physics}, S.~Jadach, G.~Passarino and R.~Pittau eds.,
CERN 2000-009 (2000).  

\bibitem{wph1} 
E.~Accomando and A.~Ballestrero, \cpc 99 1997 270 \ [hep-ph/9607317].

\bibitem{fort}
T.~Sj\"ostrand, \cpc 82 1994 74;\\
G.~J.~van~Oldenborgh, P.~J.~Franzini and A.~Borrelli, \cpc  83 1994 14 
\ [hep-ph/9402298];\\
F.~A.~Berends, R.~Pittau and R.~Kleiss, \cpc  85 1995 437 \ [hep-ph/9409326];\\
F.~Caravaglios and M.~Moretti, \pl B358 1995 332 \ [hep-ph/9507237];
F.~Caravaglios and M.~Moretti, Z. Phys. {\bf C74} (1997) 291 \ [hep-ph/9604316];\\
M.~Skrzypek, S.~Jadach, W.~Placzek, Z.~Was, \cpc 94 1996 216;\\
G.~Montagna, O.~Nicrosini and F.~Piccinini, \cpc  90 1995 141 \ [hep-ph/9506208];
G.~Charlton, G.~Montagna, O.~Nicrosini, F.~Piccinini, \cpc 99 1997 355 
\ [hep-ph/9609321]; \\
H.~Anlauf, P.~Manakos, T.~Ohl, H.~D.~Dahmen, hep-ph/9605457; \\
J.~Fujimoto, {\it et al.}, \cpc 100 1997 128
\ [hep-ph/9605312];\\
G.~Passarino, \cpc 97 1996 261 \ [hep-ph/9602302]; \\ 
D.~Bardin {\it et al.},
Nucl. Phys. Proc. Suppl. {\bf 37B} (1994) 148 \ [hep-ph/9406340];\\
D~Yu.~Bardin {\it et al.},
\cpc 104 1997 161 \ [hep-ph/9612409].


\bibitem{wweg} 
D.~Bardin {\it et al.}, {\it 'Event Generators for WW Physics'} 
[hep-ph/9709270] in Ref.~\cite{yr}, vol.2 pg. 3.

\bibitem{dpeg} 
M.~Mangano {\it et al.}, {\it 'Event Generators for Discovery Physics'}
[hep-ph/9602203]  in Ref.~\cite{yr}, vol.2 pg. 299.

\bibitem{grc4f} J.~Fujimoto {\it et al.}, \cpc 100 1997 128 \ [hep-ph/9605312].

\bibitem{u1gi} 
F.~A.~Berends and G.~B.~West, Phys. Rev. {\bf D1} (1970) 122;
Y.~Kurihara, D.~Perret-Gallix, Y.~Shimizu, Phys. Lett. {\bf B349} (1995) 367
\ [hep-ph/9412215].

\bibitem{bhf}
E.~N.~Argyres {\it et al.}, \pl  B358 1995 339 \ [hep-ph/9507216];\\
W.~Beenakker {\it et al.}, Nucl. Phys. {\bf B500} (1997) 255 \ [hep-ph/9612260];\\
U.~Baur and D.~Zeppenfeld, Phys. Rev. Lett. {\bf 75} (1995) 1002 
\ [hep-ph/9503344].

\bibitem{ifl}
E.~Accomando, A.~Ballestrero and E.~Maina, \pl B479 2000 209 \ [hep-ph/9911489].

\bibitem{efl} 
G.~Passarino, \np  B574 2000 451 \ [hep-ph/9911482]; \\
G.~Passarino, \np B578 2000 3 \ [hep-ph/0001212].

\bibitem{koralw}
S.~Jadach {\it et al.}, \cpc 140 2001 475
\ [hep-ph/0104049];\\
S.~Jadach, {\it et al.}, \cpc 119 1999 272
\ [hep-ph/9906277].

\bibitem{comphep}
A.~S.~Belyaev {\it et al.}, hep-ph/0101232;
D.~N.~Kovalenko and A.~E.~Pukhov, Nucl. Instrum. Meth. {\bf A389} (1997) 299;
A.~E.~Pukhov {\it et al.}, hep-ph/9908288;
E.~E.~Boos {\it et al.}, hep-ph/9503280.

\bibitem{grace} T.~Ishikawa {\it et al.}, KEK Report 92--19, 1993,
The GRACE Manual Ver.1.0;\\
see also H.~Tanaka, \cpc 58 1990 153; H.~Tanaka, T.~Kaneko and Y.~Shimizu,
\cpc 64 1991 149.

\bibitem{nextcalibur}
F.~A.~Berends, C.~G.~Papadopoulos and R.~Pittau, \cpc 136 2001 148 
\ [hep-ph/0011031].

\bibitem{swap} G.~Montagna {\it et al.}, \epj C20 2001 217 \ [hep-ph/0005121];

\bibitem{4f} 
M.~W.~Gr\"unewald {\it et al.}, {\it 'Four-Fermion Production in 
Electron-Positron Collisions'} [hep-ph/0005309] in Ref.~\cite{yr2k},
pg. 1.

\bibitem{dpa}
R.~G.~Stuart, Phys. Lett. {\bf B262}(1991)113;\\
A.~Aeppli, G.~J.~van Oldenborgh and D.~Wyler, Nucl. Phys. {\bf B428}(1994)126 
\ [hep-ph/9312212];\\
W.~Beenakker, F.~A.~Berends and P.~Chapovsky, Nucl. Phys. {\bf B548}(1999)3
\ [hep-ph/9811481]; \\
A.~Denner, S.~Dittmaier and M.~Roth, Nucl. Phys. {\bf B519}(1998)39 
\ [hep-ph/9710521].

\bibitem{racoonww}
A. Denner, S.~Dittmaier, M.~Roth and D.~Wackeroth, Phys. Lett. {\bf B475}
(2000)127 \ [hep-ph/9912261];\\
A.~Denner, S.~Dittmaier, M.~Roth and D.~Wackeroth, J. Phys. {\bf G26}(2000)
593 \ [hep-ph/9912290];\\
A.~Denner, S.~Dittmaier, M.~Roth and D.~Wackeroth, Eur. Phys. J. direct 
{\bf C4}(2000)1 \ [hep-ph/9912447];\\
A.~Denner, S.~Dittmaier, M.~Roth and D.~Wackeroth, Nucl. Phys. {\bf B587}
(2000)67 \ [hep-ph/0006307].


\bibitem{yfsww}
S.~Jadach {\it et al.}, \cpc 140 2001 432
\ [hep-ph/0103163];\\
S.~Jadach {\it et al.}, Phys. Lett. {\bf B417} (1998) 326 \ [hep-ph/9705429];\\
S.~Jadach {\it et al.}, Phys. Rev. {\bf D61} (2000) 113010 \ [hep-ph/9907436].

\bibitem{qedps}
J.~Fujimoto, Y.~Shimizu, T.~Munehisa, Prog. Theor. Phys. {\bf 91} (1994) 333 
\ [hep-ph/9311368]; \\
T.~Munehisa, J.~Fujimoto, Y.~Kurihara, Y.~Shimizu,
Prog. Theor. Phys. {\bf 95} (1996) 375 \ [hep-ph/9603322];\\
Y.~Kurihara {\it et al.}, Prog.Theor.Phys. {\bf 103} (2000) 1199 
\ [hep-ph/9912520];\\
Y.~Kurihara {\it et al.}, Eur. Phys. J. {\bf C20} (2001)253 \ [hep-ph/0011276].

\bibitem{boonek}
M.~Boonekamp, hep-ph/0111213.

\bibitem{circe} 
T.~Ohl, \cpc 101 1997 269 \ [hep-ph/9607454].

\bibitem{method} A.~Ballestrero and E.~Maina, \pl B350 1995 225 
\ [hep-ph/9403244].

\bibitem{phact} 
A.~Ballestrero, {\it 'PHACT 1.0 - Program for Helicity Amplitudes Calculations 
with Tau matrices'} [hep-ph/9911318] in{\it 
Proceedings of the 14th International Workshop on High Energy Physics 
and Quantum Field Theory (QFTHEP 99)}, 
 B.B.~Levchenko and V.I.~Savrin  eds. (SINP MSU Moscow), pg. 303. 

\bibitem{vegas} G.~P.~Lepage, {\it Jour. Comp. Phys.} {\bf 27} (1978) 192.

\bibitem{pythia}
T.~Sj\"ostrand {\it et al.}, \cpc 135 2001 238 \ [hep-ph/0010017];\\
T.~Sj\"ostrand, L.~L\"onnblab and S.~Mrenna, hep-ph/0108264.

\bibitem{tb}  Report on {\it QCD Event Generators}
 in ref.~\cite{yr}

\bibitem{swphasespace} Bhattacharya, J. Smith, G. Grammer, Jr. \pr D15 1977 3267

\bibitem{isr1} E.~A.~Kuraev and V.~S.~Fadin, Sov.~J.~Nucl.~Phys. {\bf 41}
 (1985) 466;\\
G.~Altarelli and G.~Martinelli, in {\it Physics at LEP}, CERN Report 86-02,
J.~Ellis and R.~Peccei eds. (CERN, Geneva 1986) pg. 47.

\bibitem{sfgg} 
W.~Beenakker, F.~A.~Berends and S.~C.~van~der~Marck, Nucl. Phys. {\bf B349}
(1991)323; \\
Y.~Kurihara {\it et al.}, Prog. Theor. Phys. {\bf 103}(2000)1199 
\ [hep-ph/9912520].

\bibitem{sfpv}
G.~Montagna {\it et al.}, Eur. Phys. J. {\bf C20}(2001)217 \ [hep-ph/0005121].

\bibitem{sfkur}
Y.~Kurihara {\it et al.}, Eur. Phys. J. {\bf C20}(2001)253 \ [hep-ph/0011276].

\bibitem{giampisr}
G.~Passarino, Nucl. Phys. {\bf B619} (2001) 313 \ [hep-ph/0108255];\\
G.~Passarino, hep-ph/0101139

\bibitem{photos} 
E. Barberio, Z. Was \cpc 79 1994 291;\\
E. Barberio, B. van Eijk and Z. Was \cpc 66 1991 115.

\bibitem{icoul1} 
D.~Bardin, W.~Beenakker and A.~Denner, \pl 317 1993 213.

\bibitem{icoul2} 
A.~P.~Chapovsky, V.~A.~Khoze, Eur. Phys. J. {\bf C9} (1999) 449 
\ [hep-ph/9902343].

\bibitem{icoul3} 
V.~S.~Fadin, V.~A.~Khoze, A.~D.~Martin and A.~Chapovsky, Phys. Rev. {\bf D52} 
(1995)1377 \ [hep-ph/9501214].
V.~S.~Fadin, V.~A.~Khoze, A.~D.~Martin, Phys. Lett. {\bf B311}(1993)311.    

\bibitem{wto} 
G.~Passarino, \cpc 97 1996 261 \ [hep-ph/9602302].

\bibitem{2loophiggs}
M.~Carena, J.~R.~Espinosa, M.~Quiros and C.~E.~M.~Wagner, Phys. Lett. 
{\bf B355}(1995)209 \ [hep-ph/9504316]. 

\bibitem{ancoup} 
M.~Bilenky, J.~L.~Kneur, F.~M.~Renard and D.~Schildknecht, \np B409 1993 
22;\\
F.~A.~Berends and A.~L.~van~Sighem, \np B454 1995 467 \ [hep-ph/9506391].

\end{thebibliography}
\end{document}